# High-resolution Modeling of Coronae and Winds in Solar-type Stars with Varying Rotation Rates. II. Stellar Winds

Yue-Hong Chen (陈悦虹)[1,2,3], Julián D. Alvarado-Gómez[3], Xin Cheng[1,2], Victor See[4], Yu Dai[1,2], Maarit J. Korpi-Lagg[5,6,7], Jörn Warnecke[5,6], Chen Xing[1,2], and Mingde Ding[2]
[1] School of Astronomy and Space Science, Nanjing University, Nanjing 210023, People's Republic of China; yh_chen@smail.nju.edu.cn, xincheng@nju.edu.cn
[2] Key Laboratory of Modern Astronomy and Astrophysics (Nanjing University), Ministry of Education, Nanjing 210023, People's Republic of China
[3] Leibniz Institute for Astrophysics Potsdam, Potsdam 14482, Germany
[4] School of Physics and Astronomy, University of Birmingham, Edgbaston, Birmingham B15 2TT, UK
[5] Department of Computer Science, Aalto University, PO Box 15400, FI-00 076 Espoo, Finland
[6] Max-Planck-Institut für Sonnensystemforschung, Justus-von-Liebig-Weg 3, D-37077 Göttingen, Germany
[7] Nordita, KTH Royal Institute of Technology & Stockholm University, Hannes Alfvéns väg 12, SE-11419 Stockholm, Sweden
*Received 2026 June 24; revised 2026 August 12; accepted 2026 September 21; published 2026 October 26*

## Abstract

Observations suggest a connection between steady-wind mass loss and coronal X-ray activity in low-mass main-sequence stars. Interpreting this connection is challenging because the wind is controlled mainly by the large-scale open field, whereas X-ray emission traces heating in small-scale closed fields often unresolved in global wind models. Here we use Space Weather Modeling Framework–Alfvén-Wave Solar Model with global convective dynamo-generated magnetic maps and solar magnetograms. We model Alfvén-wave-heated winds for four solar-type stars plus the Sun, spanning rotation rates of 1.0–23.3 times the solar rate, and magnetic field strengths of 6.0–1200 G. Our models show that faster rotation yields a more tightly wound spiral, a larger Alfvén surface, higher terminal wind speeds, and a harsher wind-pressure environment for orbiting exoplanets, different from that of the present-day Sun. We estimate the mass- and angular-momentum-loss rates and find systematic differences from Zeeman–Doppler imaging based predictions. Building on the successful reproduction of X-ray coronae in our Paper I, we obtain the first self-consistent activity–wind relation in a unified modeling framework: the mass-loss rate scales with the surface X-ray flux following a power law with an index of ∼0.67. We also reexamine magnetic braking via open-flux magnetization, finding that the effective Alfvénic lever arm depends on the magnetization parameter with a power-law index of ∼0.35. Finally, we quantify the stellar wind pressure at the orbits of several super-Earths. Together with our Paper I, the series of results shows the distinct roles of multiscale magnetic fields and provides physically grounded inputs for assessing habitable-zone space weather.

## 1. Introduction

Stellar winds correspond to magnetized plasma outflows launched from stellar coronae and guided by the large-scale open magnetic field. They continuously remove mass and angular momentum from the star and propagate throughout planetary systems (E. N. Parker 1958; E. J. Weber & L. Davis 1967; L. Mestel 1968; S. D. Kawaler 1988; J. Bouvier et al. 2014). Although other processes can dominate angular-momentum and mass evolution in different regimes (e.g., star–disk coupling at young ages; S. Matt & R. E. Pudritz 2005, or tides in close binaries; J.-P. Zahn 1977), steady winds provide the baseline for low-mass main-sequence stars. A key focus is to delineate the wind's three-dimensional (3D) structure and to quantify its dependence on stellar parameters, especially the coronal and magnetic conditions (e.g., A. A. Vidotto et al. 2011; C. Garraffo et al. 2015; J. D. Alvarado-Gómez et al. 2016a, 2016b). With wind properties predictably linked to their magnetic and coronal source, another question is how robustly such predictions carry over to rotational evolution (e.g., C. P. Johnstone et al. 2015a, 2015b) and the space-weather conditions experienced by planets, including their habitability (e.g., V. S. Airapetian et al. 2020; G. Gronoff et al. 2020).

Currently, measurements of winds remain challenging and indirect, owing to the low densities of stellar winds and the large distances from Earth to other stars (e.g., A. A. Vidotto & V. Bourrier 2017; B. E. Wood et al. 2021; K. G. Kislyakova et al. 2024). A key inference from present observations is an empirical power-law relation between the mass-loss rate and coronal X-ray flux, $\dot{M}$–$F_X^\alpha$ (e.g., B. E. Wood et al. 2002, 2005, 2014, 2021; K. G. Kislyakova et al. 2024). However, this relation is subject to significant uncertainties and observational limitations. First, the behavior at the high-activity end remains an unsolved puzzle (the putative "wind divide," e.g., B. E. Wood et al. 2021). Second, astrospheric wind measurements inherently represent spatially integrated and time-averaged properties over long periods. Interpreting this behavior is difficult in modeling because the steady wind is controlled mainly by the large-scale open field, whereas X-ray emission primarily traces plasma heating within magnetically closed structures that are frequently unresolved in observations and, consequently, are often neglected or represented only in a highly parameterized manner in global wind simulations (V. S. Airapetian et al. 2021).

Stellar winds have been investigated using analytic and semianalytic formalisms, as well as global 3D MHD simulations (e.g., E. J. Weber & L. Davis 1967; S. R. Cranmer & S. H. Saar 2011; A. A. Vidotto et al. 2011; S. P. Matt et al. 2012; O. Cohen & J. J. Drake 2014; V. Réville et al. 2015; J. D. Alvarado-Gómez et al. 2016b; A. J. Finley & S. P. Matt 2017; C. Garraffo et al. 2018; V. See et al. 2019a; J. J. Chebly et al. 2023; D. Evensberget et al. 2023). Contemporary wind models encompass a broad spectrum of driving mechanisms and parameterizations, from polytropic/thermally driven outflows to Alfvén-wave heating. Various surface magnetic fields, either idealized theoretical configurations or observed maps (e.g., from Zeeman–Doppler imaging, ZDI), are also adopted as boundary conditions. As introduced above, wind modeling places greater emphasis on the role of large-scale magnetic structures, which most directly regulate the steady wind. A robust solution is that the large-scale field and the distribution of open flux set the effective lever arm and hence the angular-momentum-loss rate. Models combined with wind–magnetic braking further support a self-regulating evolutionary picture in which magnetic energy release accelerates the winds that remove angular momentum, driving spin-down. Through the rotation–activity connection, it leads to a decline in magnetic activity and wind strength over time (J. Bouvier et al. 2014; C. Garraffo et al. 2018; A. A. Vidotto 2021). Nevertheless, from an evolutionary standpoint, ensuring the robustness of the angular-momentum-loss prescriptions employed to predict the observable spin-down of stars, a systematic evaluation of the braking law formulations across distinct acceleration regimes is required. While the scalings (e.g., S. P. Matt et al. 2012; V. Réville et al. 2015; A. J. Finley & S. P. Matt 2018; M. Shoda et al. 2020) provide a canonical description based on polytropic physics, applying them to physically distinct 3D Alfvén-wave-driven winds can serve as a valuable cross-validation framework for constraining and interpreting stellar rotational evolution.

In this work, we construct wind models that (i) are driven by Alfvén waves with the Space Weather Modeling Framework–Alfvén-Wave Solar Model (SWMF-AWSoM), and (ii) explicitly include multiscale magnetic structures in the surface boundary by using surface magnetic maps from global convective dynamo simulations (M. Viviani et al. 2018). A sequence of solar-type stars is simulated at fixed stellar mass and radius ($M_\star = 1\,M_\odot$, $R_\star = 1\,R_\odot$) over 1.0–23.3 $\Omega_\odot$ ($P_{\rm rot}$ of 25.38–1.09 days) and magnetic field strength of 6.0–1200 G. The resulting 3D solutions are used (i) to present global 3D structures of winds, (ii) to establish magnetic braking scaling relations and assess their implications for rotational evolution, and (iii) to evaluate the wind environments around solar-type stars. Our Paper I (Y.-H. Chen et al. 2025) established the first coronal and X-ray properties based on high-resolution modeling of coronae in solar-type stars. This paper, focusing on steady winds, is the second part of the same model framework. Section 2 describes the numerical methodology and Section 3 presents the main results. Section 4 discusses the implications, which is followed by the conclusions in Section 5.

## 2. Methods

### 2.1. Dynamo-generated Maps

We simulate the stellar atmospheres for the Sun and four solar-type stars (C1–C4). The surface radial magnetic field maps are input, as shown in Figure 1. For the Sun, we use two observed synoptic magnetograms from the Helioseismic and Magnetic Imager (HMI; J. Schou et al. 2012) on board the Solar Dynamics Observatory (SDO; W. D. Pesnell et al. 2012) (minimum activity phase: Carrington rotation [CR] 2223; maximum activity phase: CR 2287). For the solar-type stars, we adopt radial maps from the global convective dynamo (GCD) simulations of M. Viviani et al. (2018). These dynamo runs are constructed for solar analogs with the same fundamental parameters, namely $M_\star = 1\,M_\odot$ and $R_\star = 1\,R_\odot$, while the rotation rate is varied between cases ($\Omega$ for C1–C4 is 1.8 $\Omega_\odot$, 4.8 $\Omega_\odot$, 7.8 $\Omega_\odot$, and 23.3 $\Omega_\odot$, respectively.) Therefore, differences among our model stars are primarily driven by their different rotation periods. The dynamo runs of M. Viviani et al. (2018) that we adopt (their C1, $G^{\rm W}$, $H^{\rm a}$, and $L^{\rm a}$, corresponding to our C1–C4) show genuine cyclic variability. C1 cycles via an axisymmetric ($m = 0$) mode producing a latitudinal dynamo wave with polarity reversals ($\tau_{\rm cyc} \simeq$ 3.5 yr), as does C2 ($\tau_{\rm cyc} \simeq 2.4$ yr), which is a $\pi/2$ wedge run in which large-scale non-axisymmetric modes are suppressed by construction. C3 and C4 are instead dominated by an $m = 1$ mode (two active longitudes) whose strength oscillates in antiphase between the northern and southern hemispheres ($\tau_{\rm cyc} \simeq 7.2$ and 3.1 yr). All periods are shorter than the Sun's 11 yr cycle. Each run spans 5–16 such cycles, from which we select the extrema in surface-averaged unsigned radial field as the activity maximum and minimum. As introduced by Paper I, we apply a uniform amplitude calibration to the global convective dynamo-generated magnetograms by rescaling $B_r$ by a constant factor of 1/3. This places the resulting mean surface field strengths within the range implied by the empirical rotation–magnetism relation for unsaturated solar-type stars inferred from Zeeman broadening measurements (O. Kochukhov et al. 2020). The model takes $B_r$ at the stellar surface as input and reconstructs the full vector field via the Potential Field Source Surface (PFSS) method to initialize $B_\theta$ and $B_\phi$ in 3D space. Tables 1 and 2 list the rotation period $P$, the rotation rate $\Omega$, the mean magnetic field strength $\langle|\boldsymbol{B}|\rangle$ at $r = 1.001\,R_\odot$, and its unsigned radial component $\langle B_r\rangle$ used to set up each case.

### 2.2. Numerical Setting

We use the Alfvén Wave Solar Atmosphere Model (AWSoM; B. van der Holst et al. 2010; I. V. Sokolov et al. 2013; B. van der Holst et al. 2014; X. Meng et al. 2015; B. van der Holst et al. 2022) within the Space Weather Modeling Framework (SWMF; G. Tóth et al. 2005, 2012; T. I. Gombosi et al. 2021). AWSoM solves the extended MHD equations and includes physics-based Alfvén-wave turbulence heating and solar/stellar-wind acceleration. We first use the Solar Corona (SC) component in spherical geometry to model the stellar coronae. For completeness, we summarize here the key numerical setup (see Paper I for full details). The SC domain extends from 1 $R_\star$ to 40 $R_\star$ (80 $R_\star$ for C4), with a base angular resolution of 2°.8 refined to 0°.35 (Sun) or 0°.7 (solar-type stars) within $r < 2\,R_\star$, and a smallest radial spacing of $\Delta r \approx 0.0003$–$0.0004\,R_\odot$. The SC component contains 121 million cells for the Sun, 18 million cells for C1–C3, and 23 million cells for C4. Then, we couple the Inner Heliosphere (IH) component with SC to simulate interplanetary space. The corresponding domain is a square box, extending from $-400\,R_\odot$ to 400 $R_\odot$ in the $X$, $Y$, and $Z$ directions, and the cell size

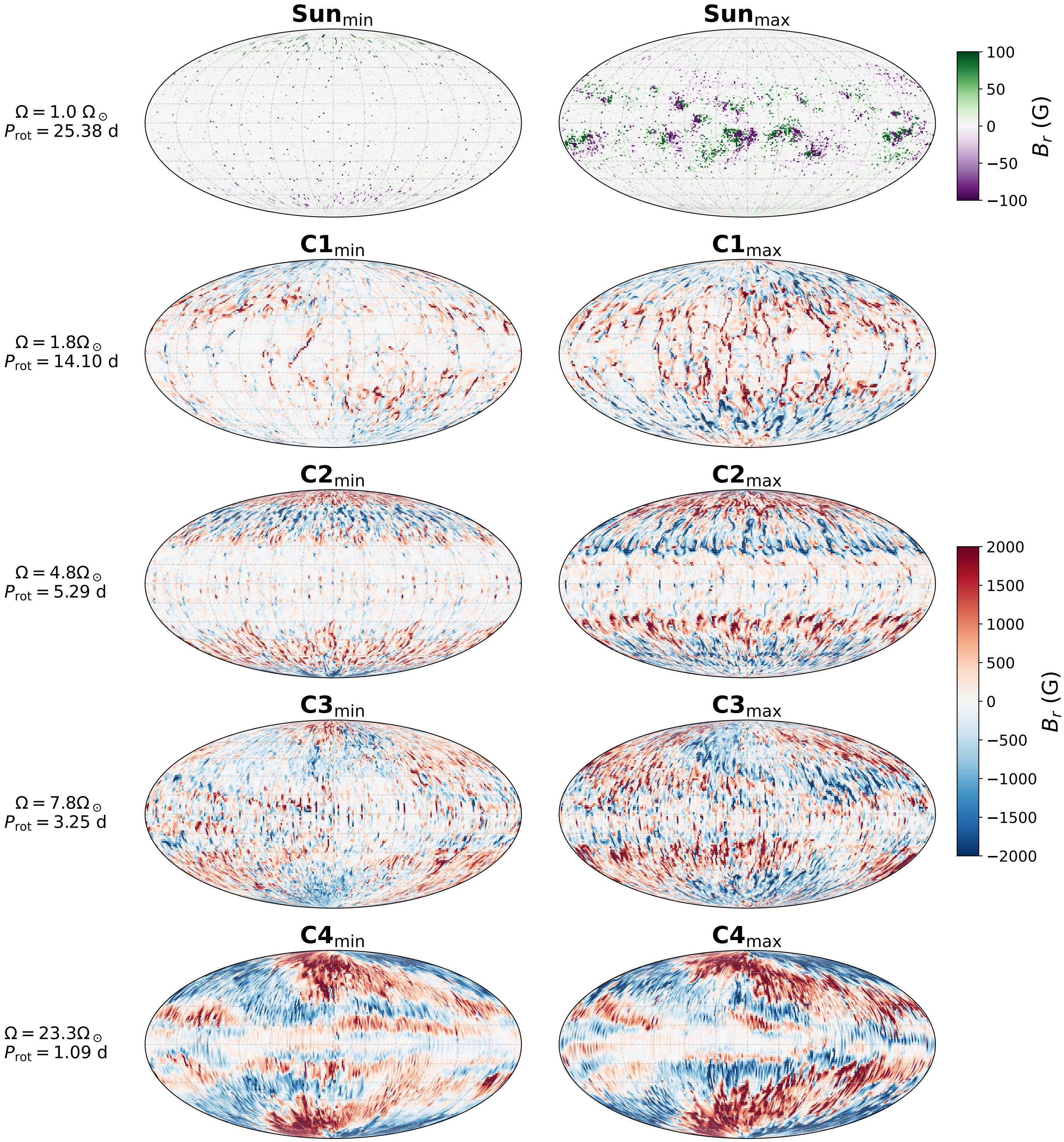

**Figure 1.** Input surface radial magnetic field ($B_r$) in a Mollweide projection. The first row shows solar radial magnetograms from SDO/HMI (left: CR2223; right: CR2287), while the remaining rows C1–C4 display dynamo-generated radial maps from M. Viviani et al. (2018). The surface radial field for C1–C4 is shown after uniformly rescaling the dynamo output by a factor of 1/3, i.e., $B_r = B_r^{\rm dyn}/3$. This scaling is chosen so that the magnetic field strength in our models is consistent with Zeeman broadening observations of young solar-type stars (O. Kochukhov et al. 2020). The left and right columns represent the stellar activity minimum and maximum phases, respectively.

($\simeq$2.8$R_\odot$) is set up to match that at the outer boundary of the SC grid. The total number of cells in IH is 56.6 million for each case. We use local time stepping (G. Tóth et al. 2012) to accelerate convergence to a steady state, which we define by requiring that, over 10,000 iterations, the X-ray luminosity changes by less than 5%, the mean coronal density and temperature at 1.5, 5, and 10 $R_\star$ change by less than 1%, and, additionally, for the wind properties as reported in this paper, the mass-loss rate $\dot{M}$ and angular-momentum-loss rate $\dot{J}$ change by less than 5% (typically within about 1% in practice) over the same window. Each case typically requires 120,000–160,000 iterations in total to reach this converged state.

**Table 1**
Summary of Simulation Parameters and Results for the Sun, C1, and C2 Cases

| Case | $\mathrm{Sun}_{\mathrm{min}}$ | $\mathrm{Sun}_{\mathrm{max}}$ | $\mathrm{C1}_{\mathrm{min}}$ | $\mathrm{C1}_{\mathrm{max}}$ | $\mathrm{C2}_{\mathrm{min}}$ | $\mathrm{C2}_{\mathrm{max}}$ |
|---|---|---|---|---|---|---|
| $P_{\mathrm{rot}}$ (day) | 25.38 | 25.38 | 14.10 | 14.10 | 5.29 | 5.29 |
| $\Omega$ ($\Omega_\odot$) | 1.0 | 1.0 | 1.8 | 1.8 | 4.8 | 4.8 |
| $Ro$[a] | 1.84 | 1.84 | 1.02 | 1.02 | 0.38 | 0.38 |
| $\langle\lvert\boldsymbol{B}\rvert\rangle$ (G) | 6.02 | 27.7 | 344 | 690 | 476 | 810 |
| $\langle B_r\rangle$ (G) | 3.58 | 15.7 | 203 | 408 | 290 | 482 |
| $S_{\mathrm{A}}/B$ (MW m$^{-2}$ T$^{-1}$) | 1 | 1 | 1.69 | 2.72 | 2.10 | 3.05 |
| $\langle R_{\mathrm{A}}\rangle_{\mathrm{geo}}$ ($R_*$) | 8.4 | 8.4 | 17.4 | 24.8 | 15.8 | 14.9 |
| $\langle R_{\mathrm{A}}\rangle_{\tau}$ ($R_*$) | 6.5 | 5.2 | 9.5 | 13.6 | 9.8 | 5.5 |
| $\dot{J}$ (erg) | $8.26 \times 10^{29}$ | $1.38 \times 10^{30}$ | $1.08 \times 10^{32}$ | $4.75 \times 10^{32}$ | $5.19 \times 10^{32}$ | $3.41 \times 10^{32}$ |
| $\dot{M}$ (g s$^{-1}$) | $1.39 \times 10^{12}$ | $3.61 \times 10^{12}$ | $4.73 \times 10^{13}$ | $1.03 \times 10^{14}$ | $8.20 \times 10^{13}$ | $1.67 \times 10^{14}$ |
| $\Phi_{\mathrm{open}}$ (Mx) | $5.24 \times 10^{22}$ | $5.39 \times 10^{22}$ | $5.1 \times 10^{23}$ | $1.46 \times 10^{24}$ | $8.87 \times 10^{23}$ | $7.75 \times 10^{23}$ |
| $f$ | 0.0046 | 0.0046 | 0.0082 | 0.0082 | 0.0219 | 0.0219 |
| $F_X$ (erg s$^{-1}$ cm$^{-2}$) | $5.31 \times 10^{3}$ | $3.60 \times 10^{4}$ | $1.69 \times 10^{6}$ | $6.44 \times 10^{6}$ | $1.66 \times 10^{6}$ | $5.98 \times 10^{6}$ |
| $p_{\mathrm{w,1\ au}}$ [nPa] | 4.40 | 2.80 | $6.62 \times 10^{1}$ | $1.81 \times 10^{2}$ | $2.01 \times 10^{2}$ | $5.25 \times 10^{2}$ |
| | [3.42, 5.63] | [1.31, $4.59 \times 10^{1}$] | [$4.02 \times 10^{1}$, $6.40 \times 10^{2}$] | [$1.39 \times 10^{2}$, $9.44 \times 10^{2}$] | [$1.37 \times 10^{2}$, $3.48 \times 10^{2}$] | [$3.93 \times 10^{2}$, $2.24 \times 10^{3}$] |
| $r_{\mathrm{M,1\ au}}$ ($R_p$) | 9.15 | 9.87 | 5.83 | 4.93 | 4.84 | 4.12 |
| | [8.79, 9.55] | [6.19, 11.20] | [3.91, 6.20] | [3.74, 5.14] | [4.42, 5.16] | [3.24, 4.33] |
| $p_{\mathrm{w,HZ1}}$ (nPa) | 8.02 | 5.90 | $1.25 \times 10^{2}$ | $3.14 \times 10^{2}$ | $3.60 \times 10^{2}$ | $1.02 \times 10^{3}$ |
| | [6.17, 9.68] | [2.84, $9.35 \times 10^{1}$] | [$7.53 \times 10^{1}$, $8.57 \times 10^{2}$] | [$2.46 \times 10^{2}$, $1.61 \times 10^{3}$] | [$2.54 \times 10^{2}$, $5.85 \times 10^{2}$] | [$7.98 \times 10^{2}$, $3.43 \times 10^{3}$] |
| $r_{\mathrm{M,HZ1}}$ ($R_p$) | 8.28 | 8.72 | 5.24 | 4.49 | 4.39 | 3.69 |
| | [8.03, 8.65] | [5.50, 9.85] | [3.72, 5.58] | [3.42, 4.68] | [4.05, 4.65] | [3.02, 3.85] |
| $p_{\mathrm{w,HZ2}}$ (nPa) | 4.92 | 3.10 | $7.45 \times 10^{1}$ | $1.99 \times 10^{2}$ | $2.29 \times 10^{2}$ | $6.18 \times 10^{2}$ |
| | [3.81, 6.21] | [1.50, $5.24 \times 10^{1}$] | [$4.53 \times 10^{1}$, $6.80 \times 10^{2}$] | [$1.54 \times 10^{2}$, $1.03 \times 10^{3}$] | [$1.53 \times 10^{2}$, $3.85 \times 10^{2}$] | [$4.46 \times 10^{2}$, $2.45 \times 10^{3}$] |
| $r_{\mathrm{M,HZ2}}$ ($R_p$) | 8.99 | 9.71 | 5.71 | 4.85 | 4.74 | 4.01 |
| | [8.64, 9.38] | [6.06, 10.95] | [3.87, 6.07] | [3.69, 5.06] | [4.34, 5.07] | [3.19, 4.24] |
| $p_{\mathrm{w,HZ3}}$ (nPa) | 1.54 | $6.48 \times 10^{-1}$ | $2.16 \times 10^{1}$ | $6.50 \times 10^{1}$ | $7.48 \times 10^{1}$ | $2.30 \times 10^{2}$ |
| | [1.15, 2.25] | [0.30, $2.09 \times 10^{1}$] | [$1.19 \times 10^{1}$, $2.30 \times 10^{2}$] | [$5.01 \times 10^{1}$, $4.75 \times 10^{2}$] | [$4.94 \times 10^{1}$, $1.26 \times 10^{2}$] | [$1.08 \times 10^{2}$, $7.22 \times 10^{2}$] |
| $r_{\mathrm{M,HZ3}}$ ($R_p$) | $1.09 \times 10^{1}$ | $1.26 \times 10^{1}$ | 7.02 | 5.84 | 5.71 | 4.73 |
| | [10.23, 11.45] | [7.06, 14.29] | [4.63, 7.60] | [4.20, 6.10] | [5.24, 6.12] | [3.91, 5.37] |
| $p_{\mathrm{w,HZ4}}$ (nPa) | 1.39 | $5.63 \times 10^{-1}$ | $1.92 \times 10^{1}$ | $5.86 \times 10^{1}$ | $6.70 \times 10^{1}$ | $2.07 \times 10^{2}$ |
| | [1.02, 2.06] | [0.26, $1.89 \times 10^{1}$] | [$1.05 \times 10^{1}$, $2.08 \times 10^{2}$] | [$4.51 \times 10^{1}$, $4.50 \times 10^{2}$] | [$4.48 \times 10^{1}$, $1.12 \times 10^{2}$] | [$9.34 \times 10^{1}$, $6.26 \times 10^{2}$] |
| $r_{\mathrm{M,HZ4}}$ ($R_p$) | $1.11 \times 10^{1}$ | $1.29 \times 10^{1}$ | 7.16 | 5.94 | 5.81 | 4.82 |
| | [10.39, 11.67] | [7.18, 14.63] | [4.71, 7.76] | [4.23, 6.21] | [5.34, 6.22] | [4.01, 5.50] |

**Notes.** Parameter definitions. $P$: stellar rotation period. $\Omega$: stellar rotation rate. $Ro$: Rossby number. $\langle\lvert\boldsymbol{B}\rvert\rangle$: mean magnetic field strength in the surface ($r = 1.001R_*$). $\langle B_r\rangle$: mean unsigned radial magnetic field strength. $S_{\mathrm{A}}/B$: Poynting-flux-related parameter in AWSoM. $\langle R_{\mathrm{A}}\rangle_{\mathrm{geo}}$: mean geometric Alfvén surface radius. $\langle R_{\mathrm{A}}\rangle_{\tau}$: mean Alfvén radius defined in Section 3.2.3. $\dot{J}$: angular-momentum-loss rate. $\dot{M}$: mass-loss rate. $\Phi_{\mathrm{open}}$: open magnetic flux at closed spherical surfaces at $R = 74\,R_*$ for C4 and at $R = 36\,R_*$ for the others. $f$: breakup fraction. $F_X$: surface X-ray flux. $p_{\mathrm{w,1au}}$, $p_{\mathrm{w,HZi}}$: median stellar-wind pressure at the 1 au, habitable zone HZ1–4 locations (0.750, 0.950, 1.676, and 1.765 au, corresponding to the recent Venus limit, the runaway greenhouse limit, the maximum greenhouse limit, and the early Mars limit using the prescription of R. K. Kopparapu et al. 2014). Values in square brackets are given as [orbital min, orbital max]. $r_{\mathrm{M,1au}}$, $r_{\mathrm{M,HZi}}$: planetary magnetopause stand-off distance (in units of planetary radius $R_p$) at the 1 au, HZ1–4 locations, as defined in Section 3.3.

[a] To compute the Rossby number, $Ro = P_{\mathrm{rot}}/\tau$, we adopt the empirically calibrated mass-dependent convective turnover time from N. J. Wright et al. (2018), $\log\tau = 2.33 - 1.50\left(\frac{M}{M_\odot}\right) + 0.31\left(\frac{M}{M_\odot}\right)^2$, where $\tau$ is in days.

**Table 2**
Same as Table 1 but for the C3, C4, and Low-resolution $C3^{LR}_{max}$ Cases

| Case | $C3_{min}$ | $C3_{max}$ | $C4_{min}$ | $C4_{max}$ | $C3^{LR}_{max}$ |
|---|---|---|---|---|---|
| $P_{rot}$ (day) | 3.25 | 3.25 | 1.09 | 1.09 | 3.25 |
| $\Omega$ ($\Omega_\odot$) | 7.8 | 7.8 | 23.3 | 23.3 | 7.8 |
| $Ro$ | 0.24 | 0.24 | 0.08 | 0.08 | 0.24 |
| $\langle\lvert\boldsymbol{B}\rvert\rangle$ (G) | 575 | 875 | 932 | 1210 | 428 |
| $\langle B_r\rangle$ (G) | 357 | 539 | 599 | 770 | 284 |
| $S_A/B$ (MW m$^{-2}$ T$^{-1}$) | 2.40 | 3.20 | 3.44 | 4.11 | 3.20 |
| $\langle R_A\rangle_{geo}$ ($R_*$) | 24.4 | 27.3 | 35.9 | 34.0 | 28.3 |
| $\langle R_A\rangle_\tau$ ($R_*$) | 15.1 | 17.0 | 15.5 | 15.4 | 17.8 |
| $\dot{J}$ (erg) | $2.23 \times 10^{33}$ | $4.47 \times 10^{33}$ | $1.61 \times 10^{34}$ | $1.81 \times 10^{34}$ | $4.44 \times 10^{33}$ |
| $\dot{M}$ (g s$^{-1}$) | $9.03 \times 10^{13}$ | $1.43 \times 10^{14}$ | $2.09 \times 10^{14}$ | $2.35 \times 10^{14}$ | $1.30 \times 10^{14}$ |
| $\Phi_{open}$ (Mx) | $1.34 \times 10^{24}$ | $2.11 \times 10^{24}$ | $2.18 \times 10^{24}$ | $2.56 \times 10^{24}$ | $2.12 \times 10^{24}$ |
| $f$ | 0.0356 | 0.0356 | 0.1064 | 0.1064 | 0.0356 |
| $F_X$ (erg s$^{-1}$ cm$^{-2}$) | $4.32 \times 10^6$ | $8.71 \times 10^6$ | $8.40 \times 10^6$ | $1.20 \times 10^7$ | ⋯ |
| $p_{w,1\ au}$ (nPa) | $1.72 \times 10^2$ [$1.50 \times 10^2$, $1.01 \times 10^3$] | $2.70 \times 10^2$ [$1.48 \times 10^2$, $2.16 \times 10^3$] | $4.48 \times 10^2$ [$2.53 \times 10^2$, $2.68 \times 10^3$] | $6.82 \times 10^2$ [$3.18 \times 10^2$, $3.99 \times 10^3$] | ⋯ |
| $r_{M,1\ au}$ ($R_p$) | 4.97 [3.70, 5.08] | 4.61 [3.26, 5.10] | 4.24 [3.14, 4.66] | 3.95 [2.94, 4.49] | ⋯ |
| $p_{w,HZ1}$ (nPa) | $3.31 \times 10^2$ [$2.70 \times 10^2$, $1.72 \times 10^3$] | $4.90 \times 10^2$ [$3.00 \times 10^2$, $3.62 \times 10^3$] | $8.14 \times 10^2$ [$4.75 \times 10^2$, $5.22 \times 10^3$] | $1.21 \times 10^3$ [$5.86 \times 10^2$, $7.62 \times 10^3$] | ⋯ |
| $r_{M,HZ1}$ ($R_p$) | 4.46 [3.39, 4.61] | 4.17 [2.99, 4.53] | 3.83 [2.81, 4.19] | 3.59 [2.64, 4.05] | ⋯ |
| $p_{w,HZ2}$ (nPa) | $1.92 \times 10^2$ [$1.66 \times 10^2$, $1.14 \times 10^3$] | $3.01 \times 10^2$ [$1.69 \times 10^2$, $2.39 \times 10^3$] | $5.01 \times 10^2$ [$2.83 \times 10^2$, $3.06 \times 10^3$] | $7.47 \times 10^2$ [$3.50 \times 10^2$, $4.48 \times 10^3$] | ⋯ |
| $r_{M,HZ2}$ ($R_p$) | 4.88 [3.63, 5.00] | 4.53 [3.20, 4.98] | 4.16 [3.08, 4.57] | 3.89 [2.89, 4.41] | ⋯ |
| $p_{w,HZ3}$ (nPa) | $5.66 \times 10^1$ [$4.34 \times 10^1$, $3.83 \times 10^2$] | $8.27 \times 10^1$ [$4.26 \times 10^1$, $8.78 \times 10^2$] | $1.56 \times 10^2$ [$8.26 \times 10^1$, $8.64 \times 10^2$] | $2.41 \times 10^2$ [$7.70 \times 10^1$, $1.30 \times 10^3$] | ⋯ |
| $r_{M,HZ3}$ ($R_p$) | 5.98 [4.35, 6.25] | 5.61 [3.79, 6.27] | 5.05 [3.80, 5.61] | 4.70 [3.55, 5.68] | ⋯ |
| $p_{w,HZ4}$ (nPa) | $5.05 \times 10^1$ [$3.81 \times 10^1$, $3.41 \times 10^2$] | $7.22 \times 10^1$ [$3.76 \times 10^1$, $8.17 \times 10^2$] | $1.40 \times 10^2$ [$7.28 \times 10^1$, $7.54 \times 10^2$] | $1.36 \times 10^2$ [$6.52 \times 10^1$, $1.00 \times 10^3$] | ⋯ |
| $r_{M,HZ4}$ ($R_p$) | 6.09 [4.43, 6.39] | 5.74 [3.83, 6.40] | 5.14 [3.88, 5.73] | 5.16 [3.70, 5.84] | ⋯ |

**Note.** See Table 1 for parameter definitions. The $C3^{LR}_{max}$ case is from Appendix A, where we truncate the spherical-harmonic coefficient of the surface map to $l_{max} = 5$ for $C3_{max}$.

For the solar cases, we prioritize the well-validated standard parameter settings from the AWSoM community. For stellar cases, as introduced in Paper I, we set the Poynting flux based on the relation between the X-ray emission and the expected input energy. With this setup, the model successfully reproduces the observed properties of the corona. Tables 1 and 2 list the Alfvén-wave Poynting-flux parameter $S_A/B$. Since the Poynting flux largely controls the resulting coronal heating and wind acceleration, we briefly recap the prescriptions adopted in Paper I. In AWSoM, the wave energy input is specified by the ratio $S_A/B$ at the inner boundary. For the Sun this ratio is fixed by the standard, well-validated calibration, $S_A/B = 1.0$ MW m$^{-2}$ T$^{-1}$ (I. V. Sokolov et al. 2013; N. Sachdeva et al. 2019, 2021, 2023; Z. Huang et al. 2023), but for other stars there is no established prescription (e.g., S. R. Cranmer & S. H. Saar 2011). Paper I therefore adopted a simplified scaling to guide this choice. Writing the Poynting flux as $S_A = \varepsilon_{wave} V_A$, with the wave energy density $\varepsilon_{wave} \propto \rho\, v'^2$ and the Alfvén speed $V_A = B/\sqrt{\mu_0 \rho}$, gives $S_A/B \propto v'^2 \rho^{1/2}$. Assuming comparable velocity fluctuations $v'$ at the base of the corona among solar-type stars (S. Boro Saikia et al. 2023), and using $L_X \propto n^2$ for optically thin coronal emission, this yields $\frac{(S_A/B)_\star}{(S_A/B)_\odot} = \left(\frac{\rho_\star}{\rho_\odot}\right)^{1/2} \simeq \left(\frac{L_{X,\star}}{L_{X,\odot}}\right)^{1/4}$, so that $S_A/B$ is determined once $L_X$ is specified. Because $L_X$ has not been measured for most of the stars in the sample, Paper I assumed it a priori from the empirical relation of O. Kochukhov et al. (2020), $\log(L_X/L_{bol}) = 2.7 \log\langle\lvert\boldsymbol{B}\rvert\rangle - 12.1$.[8] The X-ray luminosities synthesized from the resulting coronal models do not exactly reproduce the assumed values, but they remain within the observed scatter of the $L_X$–$\langle|\boldsymbol{B}|\rangle$ relation for solar-type stars. Iterating on $S_A/B$ until the two converge would be more self-consistent, but is not warranted here given that the residual differences are smaller than the scatter of the empirical relation itself.

[8] We stress that the Sun does not enter this chain: the solar $S_A/B$ is known a priori from the standard AWSoM calibration and serves as the reference point, so it is never inferred from an $S_A/B$–$\langle\lvert\boldsymbol{B}\rvert\rangle$ or $L_X$–$\langle\lvert\boldsymbol{B}\rvert\rangle$ relation. This distinction matters because the $L_X \propto \lvert\boldsymbol{B}\rvert^{2.7}$ relation of O. Kochukhov et al. (2020) is fitted over a narrow range of active young suns, and extrapolating such a steep power law down to solar-strength fields severely underestimates the solar $L_X$.

### 2.3. Key Stellar-wind Properties

We characterize the wind by the mass-loss rate $\dot{M}$ and the angular-momentum-loss rate $\dot{J}$, obtained by integrating the corresponding fluxes over a closed spherical surface $S$ of radius $r$ centered on the star, with outward unit normal $\boldsymbol{n}$ and area element $dA$:

$$\dot{M} = \oint_S \rho\,(\boldsymbol{u}\cdot\boldsymbol{n})\;dA, \tag{1}$$

$$\dot{J}_z = \oint_S \hat{z}\cdot[\,\boldsymbol{r}\times(\mathsf{T}\cdot\boldsymbol{n})\,]\,dA, \tag{2}$$

where $\rho$, $\boldsymbol{u}$, and $\boldsymbol{B}$ are the plasma density, velocity, and magnetic field, $p$ is the gas pressure, $\boldsymbol{r}$ is the position vector measured from the stellar center, $\hat{z}$ is the unit vector along the stellar rotation axis, and

$$\mathsf{T} = \rho\,\boldsymbol{u}\otimes\boldsymbol{u} + \left(p + \frac{B^2}{8\pi}\right)\mathsf{I} - \frac{\boldsymbol{B}\otimes\boldsymbol{B}}{4\pi} \tag{3}$$

is the total (Reynolds plus thermal plus Maxwell) stress tensor in Gaussian units, with $\mathsf{I}$ the unit tensor, so that $(\mathsf{T}\cdot\boldsymbol{n})_i = T_{ij}n_j$ is the force per unit area transmitted across $S$ (see L. Mestel & C. S. Selley 1970; A. A. Vidotto et al. 2014 for the full derivation). Unlike A. A. Vidotto et al. (2014), we evaluate **u** in the inertial frame. Positive values of $\dot{M}$ and $\dot{J}$ indicate outward fluxes, meaning that the star is losing mass and angular momentum. In Tables 1 and 2, the values of $\dot{M}$ and $\dot{J}$ are actually the absolute values. The integration surface $S$ is placed at $R = 74\,R_*$ for C4 and $R = 36\,R_*$ for the remaining cases, which is beyond the Alfvén surface so that the stellar winds have essentially completed their acceleration there, while remaining a bit away from the outer boundary of SC domain to minimize contamination from the boundary condition.

## 3. Results

### 3.1. Three-dimensional Structure of Stellar Winds

Figure 2 shows the global wind morphology. Rapid rotation produces more strongly twisted field lines, which lead to more spiral-shaped magnetic fields. This geometry affects the propagation of the ejected structures and energetic particles.

Figure 2 also shows the corresponding Alfvén surface (AS) for all cases. The AS is defined by $M_A = |\boldsymbol{u}|/u_A = 1$, where $u_A = B/\sqrt{4\pi\rho}$ is the local Alfvén speed, with $B$ and $\rho$ the magnetic field strength and plasma mass density, respectively. In the solar minimum case, the AS is roughly symmetric about the polar axis with a geometric-mean Alfvén radius $\langle R_A\rangle_{geo}$ of $8.45\,R_\odot$. In contrast, the solar maximum case exhibits a more irregular AS due to the presence of more active regions, with $\langle R_A\rangle_{geo} = 8.39\,R_\odot$. These values are smaller than the Parker Solar Probe (PSP)–based in-situ extrapolation estimates by S. R. Cranmer et al. (2023), who found a median $r_A = 13.4\,R_\odot$ with a standard deviation of $5.4\,R_\odot$. A more rigorous calibration against the most recent PSP measurements would likely yield improved quantitative agreement (e.g., B. van der Holst et al. 2022; Z. Huang et al. 2023). However, such an effort lies beyond the present study, which is primarily focused on a comparative analysis of the behavior across different stellar cases. In the intermediate rotation regime (C1 and C2), the polar magnetic field takes on a unipolar configuration, so that the polar regions tend to be threaded by open field lines, giving rise to fast polar wind streams and large-scale Alfvén surfaces reminiscent of those observed during solar minimum, and thus to the marked latitudinal inhomogeneity. In the most rapidly rotating cases (C3 and C4), by contrast, the polar magnetic field becomes structurally complex, and the latitudinal distribution of magnetic activity is correspondingly more uniform. The rapidly rotating cases also display substantially stronger variability in the azimuthal direction.

Tables 1 and 2 list $\langle R_A\rangle_{geo}$ for each of all cases. Relative to the solar case, the solar-type stars exhibit more extended AS. The $\langle R_A\rangle_{geo}$ is a result of the competition between the wind speed $u$ and the local Alfvén speed $u_A$. The stronger magnetic fields inject more magnetic energy into the outflow and enhance the acceleration of winds, while they also raise $u_A$. The increase in $u_A$ might outweigh the increase in $u$, shifting the $M_A = 1$ surface outward and yielding a larger AS. Among all the solar-type stars, $C4_{min}$ attains the largest value, $\langle R_A\rangle_{geo} = 35.9\,R_*$, implying the most extended inner corona, consistent with our results in Paper I.

We characterized stellar-wind speed $u$ across a spherical shell near the outer computational boundary ($r = 74\,R_\star$ for C4 and $r = 36\,R_\star$ for the others, to avoid any potential influence from the outer boundary conditions while ensuring that the wind speeds have reached their terminal values). Figure 3 presents their distributions using probability density plots. All stars except for the Sun demonstrate greater mean $u$ during the magnetic maximum phase compared to the minimum phase, likely because the contrast in magnetic field strength between the two phases is more pronounced in these stars than in the Sun. The $u$ can reach 1000 km $s^{-1}$ for all of our solar-type stars, and at 1 au it remains comparable. The speeds are substantially higher than the wind speeds commonly assumed in astrospheric Ly$\alpha$ absorption analyses (typically ~400 km $s^{-1}$; B. E. Wood et al. 2002, 2005, 2014, 2021). Because the inferred mass-loss rate for a given Ly$\alpha$ absorption signature scales as $\dot{M} \propto p_w/u$ at fixed observed absorption (fixed $p_w$), a factor of 2 underestimate in $u$ translates into a factor of 2 overestimate in $\dot{M}$. This suggests that Ly$\alpha$-derived $\dot{M}$ values may carry a systematic uncertainty tied to the assumed wind speed, particularly for young, rapidly rotating, or magnetically active stars whose winds are expected to depart significantly from solar-like conditions, as also noted by B. E. Wood et al. (2002).

The bimodal latitudinal distribution of the solar wind speed during $Sun_{min}$ arises because the poloidal component of the magnetic field reaches its maximum strength at this phase of the cycle. The dominant poloidal field drives fast solar wind streams at high latitudes, while the slow solar wind associated with the toroidal component is comparatively weak, resulting in the characteristic bimodality seen at $Sun_{min}$. In cases of moderate rotations (C1 and C2), the poloidal component remains strong throughout the cycle, and, consequently, the bimodal distribution persists at all times. This suggests that, in this rotation regime, poloidal field generation by the stellar dynamo dominates over toroidal field generation. In cases of faster rotations (C3 and C4), the toroidal component regains dominance, and the bimodality disappears. We interpret this as the signature of a further dynamo transition, in which toroidal field generation once again becomes more efficient throughout the entire activity cycle.

### 3.2. Global Losses and Controlling Parameters

#### 3.2.1. Stellar-wind Mass- and Angular-momentum-loss Rates

Using the methods introduced in Section 2.3, we calculate the stellar mass-loss rate $\dot{M}$ and angular-momentum-loss rate $\dot{J}$ from our 3D wind solutions, which are listed in Tables 1 and 2. For the Sun, we derive a mass-loss rate of $1.39 \times 10^{12}\,\mathrm{g\,s^{-1}}$ in the minimum case and $3.61 \times 10^{12}\,\mathrm{g\,s^{-1}}$ in the maximum case, in good agreement with the observationally inferred solar value $\sim 1.3 \times 10^{12}\,\mathrm{g\,s^{-1}}$ (e.g., O. Cohen 2011). They are also comparable with the empirically reconstructed solar-cycle variation of $\dot{M}$ from $\sim 1.2 \times 10^{12}\,\mathrm{g\,s^{-1}}$ at minimum to $\sim 2.0 \times 10^{12}\,\mathrm{g\,s^{-1}}$ at maximum (Y.-M. Wang 1998). Notably, our models reproduce the positive correlation between mass-loss rate and solar activity level. We find an angular-momentum-loss rate of $8.26 \times 10^{29}$ erg in the minimum case and $1.28 \times 10^{30}$ erg in the maximum case, also consistent with the current solar estimate: $\sim(0.3\text{–}6) \times 10^{30}$ erg (A. J. Finley et al. 2018, 2019).

In Figure 4, we also present our results in terms of the Rossby number ($R$o) and the rotation rate ($\Omega$). Additionally, we compare against representative studies using several wind models:

(i) A model by O. Cohen & J. J. Drake (2014) that employed a Wang–Sheeley–Arge-type semiempirical energy source term, exploring a range of dipole field strengths, rotation periods, and coronal base densities to derive scaling laws for stellar mass- and angular-momentum-loss rates. We evaluate $\dot{M}$ and $\dot{J}$ by applying their scaling laws, details introduced in Appendix B.

(ii) AWSoM models (e.g., D. Evensberget et al. 2021, 2022, 2023 and J. J. Chebly et al. 2023). For D. Evensberget et al. (2021, 2022, 2023), we use only the results computed from the original ZDI magnetic maps, and exclude those based on maps in which the magnetic field strength was artificially increased by a factor of 5. Note that J. J. Chebly et al. (2023) used a different function to calculate the $\dot{J}$, so we do not show their $\dot{J}$ results here.

(iii) The Alfvén-wave-driven mass-loss rate prescription by S. R. Cranmer & S. H. Saar (2011).

(iv) The angular-momentum-loss rates derived from a semiempirical magnetic braking law constrained by observations in S. P. Matt et al. (2015).

For (iii) and (iv), we use the tabulated values from V. See et al. (2019a), who applied these two methods to a sample of real stars, and the comparison is restricted to stars with masses and radii within [0.7, 1.3] $M_\odot$ and [0.7, 1.3] $R_\odot$, respectively.

The comparison with previous results emphasizes the following. First, the mass- and angular-momentum-loss rates inferred from different wind models span roughly 2–3 orders of magnitude at any given rotation rate or Rossby number, which summarizes the current state of various stellar-wind modeling. While systematic differences between modeling frameworks remain substantial, within each model, the rotation rate or Rossby number is clearly not the sole parameter governing the loss rates; the differences in stellar mass, radius, magnetic field, and the activity cycle all contribute to the scatter. In our cases, since the stellar radius and mass are fixed, the scatter is primarily caused by the variations in the dynamo-generated magnetic field and activity level. The scaling relations of O. Cohen & J. J. Drake (2014) are derived under the assumption of a purely dipolar field. Indeed, they also note that, when applied to a realistic solar magnetic field, their relations can deviate by up to a factor of ∼3. For our more complex field configurations, we likewise find nonnegligible deviations, consistent with the picture in which the magnetic field topology is a key factor in modulating the mass- and angular-momentum-loss rates. Second, if we restrict the comparison to studies that also employ the AWSoM model (i.e., category (ii)), our inferred loss rates are systematically higher than theirs by roughly an order of magnitude, probably because of the differences in energy input. Studies from category (ii) typically use ZDI-reconstructed surface magnetic maps as input. These maps systematically underestimate the magnetic field strengths by factors of a few (T. Hackman et al. 2024), typically due to inadequate spatial resolution and cancellation effects in observations (L. T. Lehmann et al. 2019). This can naturally give rise to such a dispersion. Another important difference is the adopted Alfvén-wave Poynting-flux input: other AWSoM-based studies typically use a solar-calibrated Poynting flux, whereas we scale the injected wave energy self-consistently with stellar X-ray luminosity, $L_X$ (see our Paper I).

Motivated by the magnetic braking (E. J. Weber & L. Davis 1967; L. Mestel 1968), we expect the wind torque to scale with the stellar rotation rate. We perform a linear fit in log–log space to $\Omega$ and $\dot{J}/R_*^2$, as shown in the third panel in Figure 4, yielding

$$\log_{10}\left(\frac{\dot{J}}{4\pi R_*^2}\right) = (2.74 \pm 0.97)\log_{10}(\Omega) + (23.20 \pm 4.79). \tag{4}$$

A comparison of the derived exponent 2.74 ± 0.97 with the scaling relations obtained in previous theoretical and observational studies will be discussed in Section 4.3.

#### 3.2.2. Mass-loss Rate versus X-Ray Flux

Stellar winds are launched from the stellar coronae, so a close connection between coronal properties and the wind parameters is expected. In particular, observations suggest that the mass-loss rate per unit surface area $\dot{M}/R_*^2$ increases with the X-ray flux over stellar surface $F_X$ (albeit with significant scatter) (e.g., B. E. Wood et al. 2002, 2005, 2014, 2021; K. G. Kislyakova et al. 2024).

Here, we test this relation with our simulations, as shown by Figure 5.[9] Our results yield a similar trend, with

$$\dot{M} \propto F_X^{0.67}, \tag{5}$$

while the observational samples follow the scaling relation $\dot{M} \propto F_X^{0.77}$. Our power-law index is also close to 0.82 estimated by T. K. Suzuki et al. (2013), where they conducted 1D MHD simulations of Alfvén-wave-driven outflows in an open flux tube. However, their X-ray flux was not computed self-consistently; they adopted the radiation from gas at $T \geqslant 2 \times 10^4$ K in the open-tube region, scaled by an empirical factor $c_r = 2$ chosen to match the observed distribution of B. E. Wood et al. (2005). M. Shoda et al. (2020) also discussed this relation, but they instead invoked the empirical activity–rotation relation of N. J. Wright et al. (2011) to convert their prediction into an $F_X$ for each star. In our 3D simulations, both open and closed regions are modeled self-consistently, enabling a more direct estimate of the coronal X-ray output (see Papar I) and a self-consistent calibration of the wind–corona relation for the first time. We discuss it further in Section 4.1. It is worth noting that excluding the solar data points from the fit yields a trend essentially consistent with that obtained when it is included, suggesting that our conclusion is reasonably robust in this regard. Nevertheless, a more reliable determination of this trend would require additional simulated points in the solar-like rotation regime.

[9] $\dot{M}_\odot = 1.3 \times 10^{12}\,\mathrm{g\,s^{-1}}$ or $2 \times 10^{-14}\,M_\odot\,\mathrm{yr^{-1}}$.

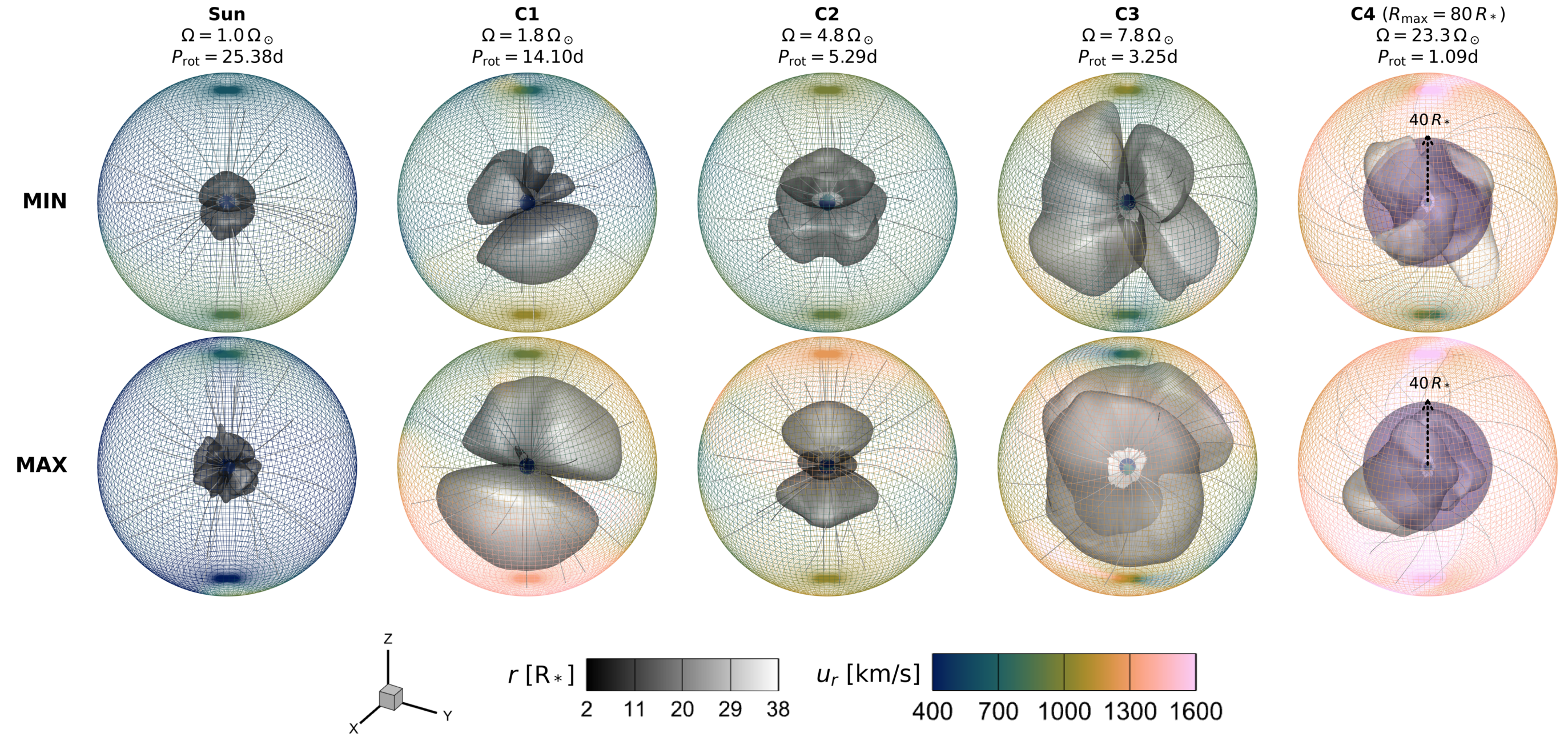

**Figure 2.** Stellar-wind structure and the Alfvén surface (AS). The spherical grids illustrate the extent of the computational domain (40 $R_*$ for the Sun and C1–C3, 80 $R_*$ for C4). The transparent spherical surfaces in the C4 cases mark the location of 40 $R_*$. Colors in the grids indicate terminal wind speeds. The gray-colored isosurfaces mark the AS. The grayscale shading on the isosurfaces indicates the radial distance from the stellar center. The lines show magnetic field lines.

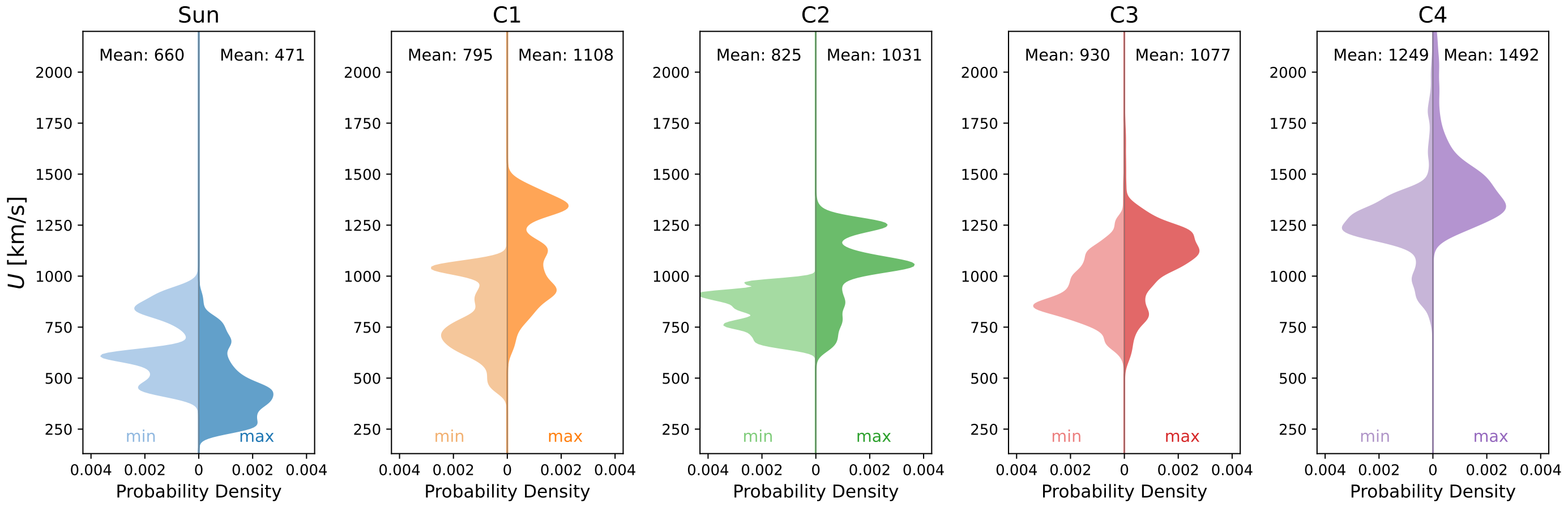

**Figure 3.** Probability density plots of the wind speed ($u$) near the outer boundary in each model ($r = 74\ R_\star$ for C4 and $r = 36\ R_\star$ for the others).

#### *3.2.3. Torque-defined Alfvén Radius versus Open Magnetic Flux*

Magnetic braking by magnetized stellar winds is commonly described as the wind extracting angular momentum as if the outflow corotates with the star out to an effective Alfvénic lever arm (E. J. Weber & L. Davis 1967). In 3D winds, the local Alfvén radius varies between field lines, but the net torque can always be expressed in terms of a torque-defined effective Alfvén radius: $\frac{\langle R_A \rangle_\tau}{R_*} = \left( \frac{j}{\dot{M}\ \Omega_*\ R_*^2} \right)^{1/2}$, where the $\langle R_A \rangle_\tau$ does not necessarily coincide with a purely geometric-mean Alfvén radius ($\langle R_A \rangle_{\rm geo}$). To obtain a simulation-calibrated prescription for the braking efficiency, it is also useful to introduce a dimensionless wind magnetization parameter that measures the relative importance of magnetic stresses compared to wind inertia. Following S. P. Matt et al. (2012) and V. Réville et al. (2015), the magnetic control of the outflow is characterized by using a dimensionless quantity $\Upsilon_{\rm open} \equiv \frac{\Phi_{\rm open}^2}{R_*^2\,\dot{M}_w\,v_{\rm esc}}$, where $v_{\rm esc} = \sqrt{\frac{2GM_*}{R_*}}$ and $\Phi_{\rm open} \equiv \oint_{Sr} |\boldsymbol{B}\cdot d\boldsymbol{S}|$. $Sr$ is a closed spherical surface of radius $r$ chosen sufficiently large so that the magnetic field is purely open there. Here we adopt the same spherical surface used in our earlier calculations of $\dot{M}$ and $\dot{J}$.

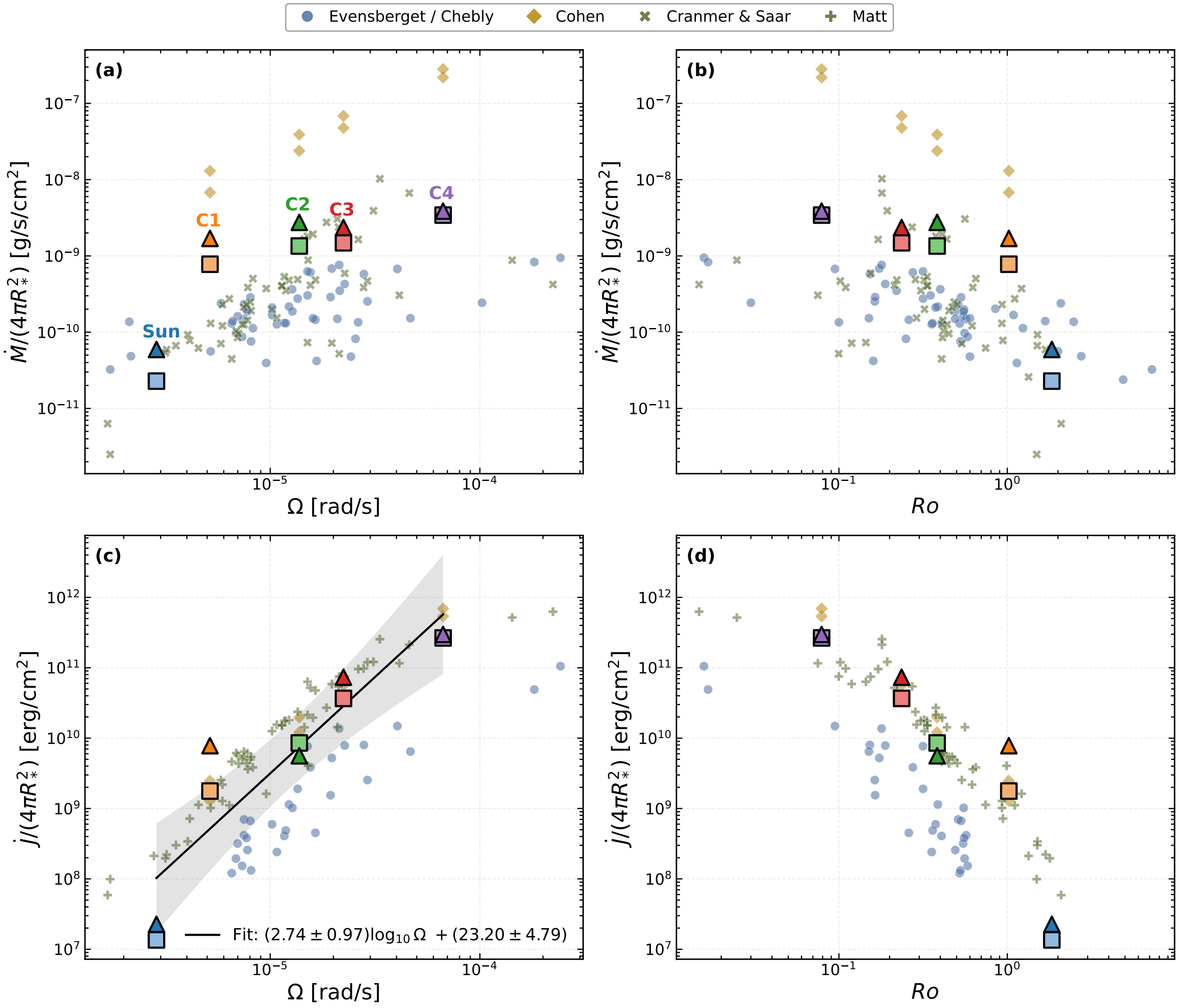

**Figure 4.** Stellar-wind mass-loss rate and angular-momentum-loss rate per unit area. Our results are plotted as triangles (at the maximum) and squares (at the minimum). The blue, orange, green, red, and purple colors correspond to the Sun and C1–C4 cases, respectively. The diamonds, circles, crosses, and plus signs represent results calculated by or collected from (i) O. Cohen & J. J. Drake (2014), (ii) D. Evensberget et al. (2021, 2022, 2023) and J. J. Chebly et al. (2023), (iii) S. R. Cranmer & S. H. Saar (2011), and (iv) S. P. Matt et al. (2015). In panel (c), we perform a least-squares linear fit in log–log space to our results, yielding the coefficient of determination $R^2 = 0.84$. The best-fit relation is shown by a black solid line, with the shaded region indicating the 95% confidence interval.

Over a broad range of parameter space, an empirical power-law relationship has been suggested to exist between the corresponding dimensionless quantities (V. Réville et al. 2015, their Equation (18)):

$$\frac{\langle R_{\rm A}\rangle}{R_*} = K_1\left(\frac{\Upsilon_{\rm open}}{\sqrt{1 + f^2/K_2^2}}\right)^m, \tag{6}$$

where $f = \frac{\Omega_* R_*^{3/2}}{\sqrt{GM_*}}$ is the breakup fraction to characterize the rotation, especially in the regime of rapid rotation where magneto-centrifugal effects significantly enhance the wind outflow velocity. When fitting the above equation, we find that the best-fit solution drives $K_2$ to very large values (e.g., $K_2 \to \infty$), so that the model effectively degenerates to the no-rotation form:

$$\frac{\langle R_{\rm A}\rangle_\tau}{R_*} = 0.29(\Upsilon_{\rm open})^{0.35}, \tag{7}$$

as shown by Figure 6.

We noticed an outlier, C2max (the green triangle in Figure 6). Although C2max and C3max both exhibit very strong surface magnetic fields (∼8–9 × $10^2$ G), C2max has a lower open magnetic flux $\Phi_{\rm open}$ and a lower magnetization parameter $\Upsilon_{\rm open}$. Its $\dot{J}$ is also about 1 order of magnitude smaller than that of C3max, yielding a smaller effective Alfvén radius $\langle R_{\rm A}\rangle_\tau$.

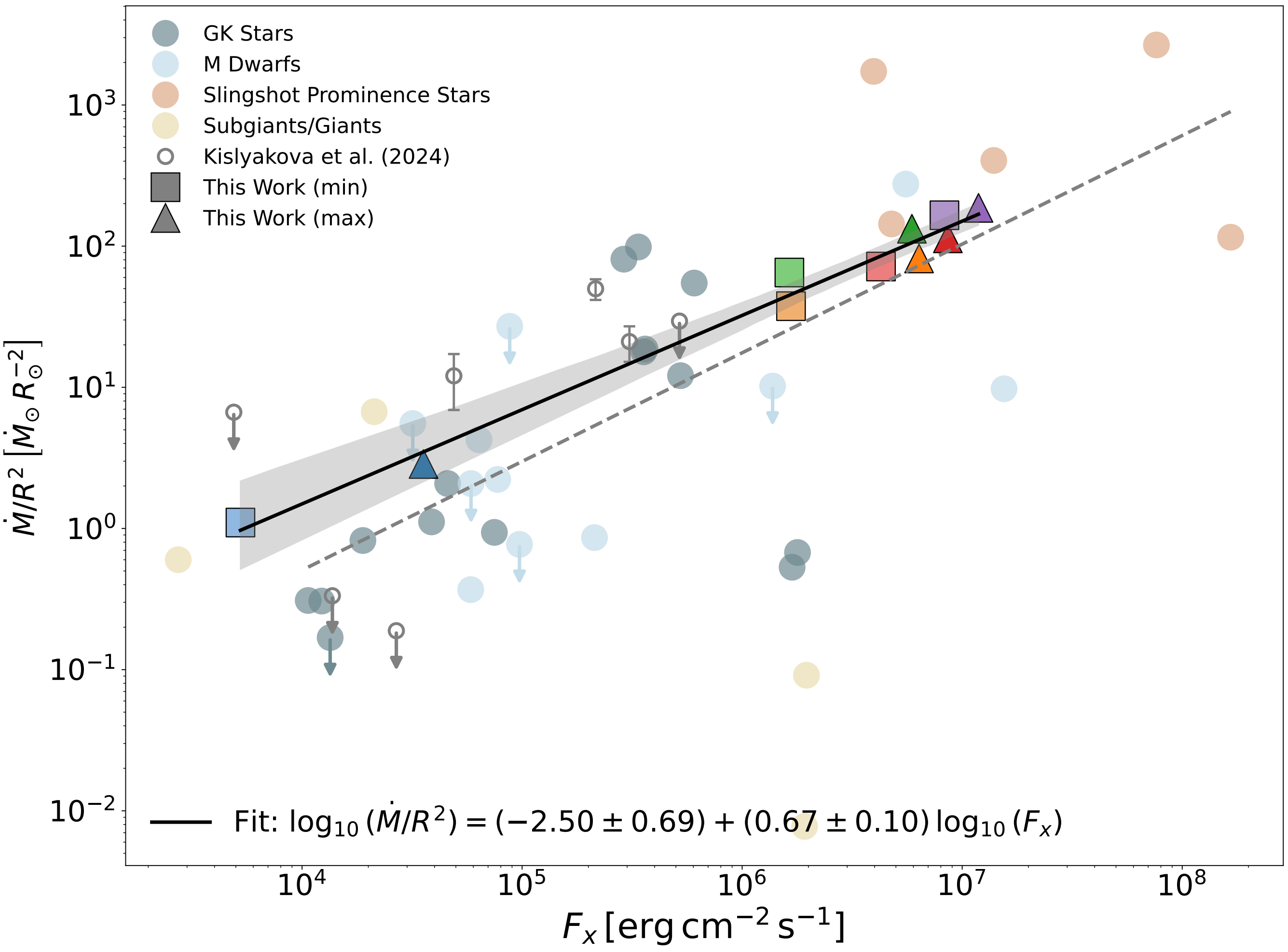

**Figure 5.** Stellar mass-loss rates as a function of surface X-ray flux ($F_X$). The diamond and triangle symbols are our minimum and maximum cases. The blue, orange, green, red, and purple colors correspond to the Sun and C1–C4 cases, respectively. The black solid line shows the best-fit power law in log–log scale, yielding $\log_{10}\dot{M}/R_*^2 \propto 0.67\log_{10}(F_X)$, and the shaded gray region indicates the 95% confidence of the fit. The coefficient of determination is $R^2 = 0.98$. The filled circular symbols are taken from Table 3 of B. E. Wood et al. (2021), which mostly come from astrospheric Ly$\alpha$ absorption detections, compiled together with slingshot-prominence wind measurements from M. Jardine & A. Collier Cameron (2019) and one Ly$\alpha$ absorption measurement (A. A. Vidotto & V. Bourrier 2017). The gray dashed line represents a fit to the combined observational data excluding subgiants and giants, resulting in $\dot{M} \propto F_X^{0.77}$. We also include mass-loss rate estimates (gray open circles) from Table 2 of K. G. Kislyakova et al. (2024), which are derived from charge-exchange induced X-ray emission.

From Figure 2, we also see that the Alfvén surface is more compact in C2$_{\rm max}$. We will discuss it further in Section 4.2.

### 3.3. Wind Pressure on Habitable Zones

The total wind pressure includes thermal, magnetic, and ram components: $p_{\rm w} = p + \frac{|\boldsymbol{B}|^2}{8\pi} + \rho\,|\boldsymbol{u} + \boldsymbol{v}|^2$, where $p$ is the thermal pressure, $\boldsymbol{B}$ is the magnetic field, $\rho$ is the plasma density, $\boldsymbol{u}$ is the wind velocity, and $\boldsymbol{v}$ is the orbital velocity of the planet. In our models, $\rho|\boldsymbol{u}|^2$ dominates $p_{\rm w}$ throughout the habitable zone (HZ). It combines thermal and magnetic contributions, $p + |\boldsymbol{B}|^2/8\pi$, amounting to less than 6% of $\rho\,|\boldsymbol{u}|^2$ everywhere within the HZ. In the following, we thus approximate the total wind pressure as $p_{\rm w} \simeq \rho\,|\boldsymbol{u}|^2$.

Figure 7 shows the spatial distribution of $p_{\rm w}$. It is visualized via direct volume rendering with `PyVista` (B. Sullivan & A. Kaszynski 2019), using the perceptually uniform `Lipari` and `viridis` color maps (F. Crameri et al. 2020). We also compute the orbit-median $p_{\rm w,1\,au}$, as summarized in Tables 1 and 2, with the minimum (−) and maximum (+) values listed. For our adopted parameters (star: 1 solar mass, planet: 1 earth mass), the conservative HZ extends from the runaway greenhouse limit at ~0.950 au (HZ2) to the maximum greenhouse limit at ~1.676 au (HZ3), while the optimistic HZ extends from the recent Venus limit at ~0.750 au (HZ1) to the early Mars limit at ~1.765 au (HZ4) (R. K. Kopparapu et al. 2013, 2014). The $p_{\rm w}$ values within these boundaries are also summarized in Table B1.

For the present-day Sun, we recover median values of $p_{\rm w,1\,au} \sim 3$ nPa, which is a bit higher than the typical near-Earth solar wind values (~2 nPa; J. H. King & N. E. Papitashvili 2005). N. Sachdeva et al. (2019) report that AWSoM systematically overestimates proton density at 1 au, which is consistent with what we see. Compared to the solar case, the active solar-type cases (C1–C4) exhibit systematically higher wind pressures, with $p_{\rm w,1\,au}$ increasing by ~1–2 orders of magnitude across our cases, indicating a substantially more extreme wind environment at Earth-like orbital distances. Within the HZ, $p_{\rm w,1\,au}$ decreases monotonically with orbital distance. The bracketed $[p_{\rm w}^-, p_{\rm w}^+]$ values in Table B1 further show that $p_{\rm w}$ can vary by up to a factor of ~10 along a fixed circular orbit, implying that the wind forcing there is phase dependent. We will further apply our wind-pressure model to real planetary systems in Section 4.4.

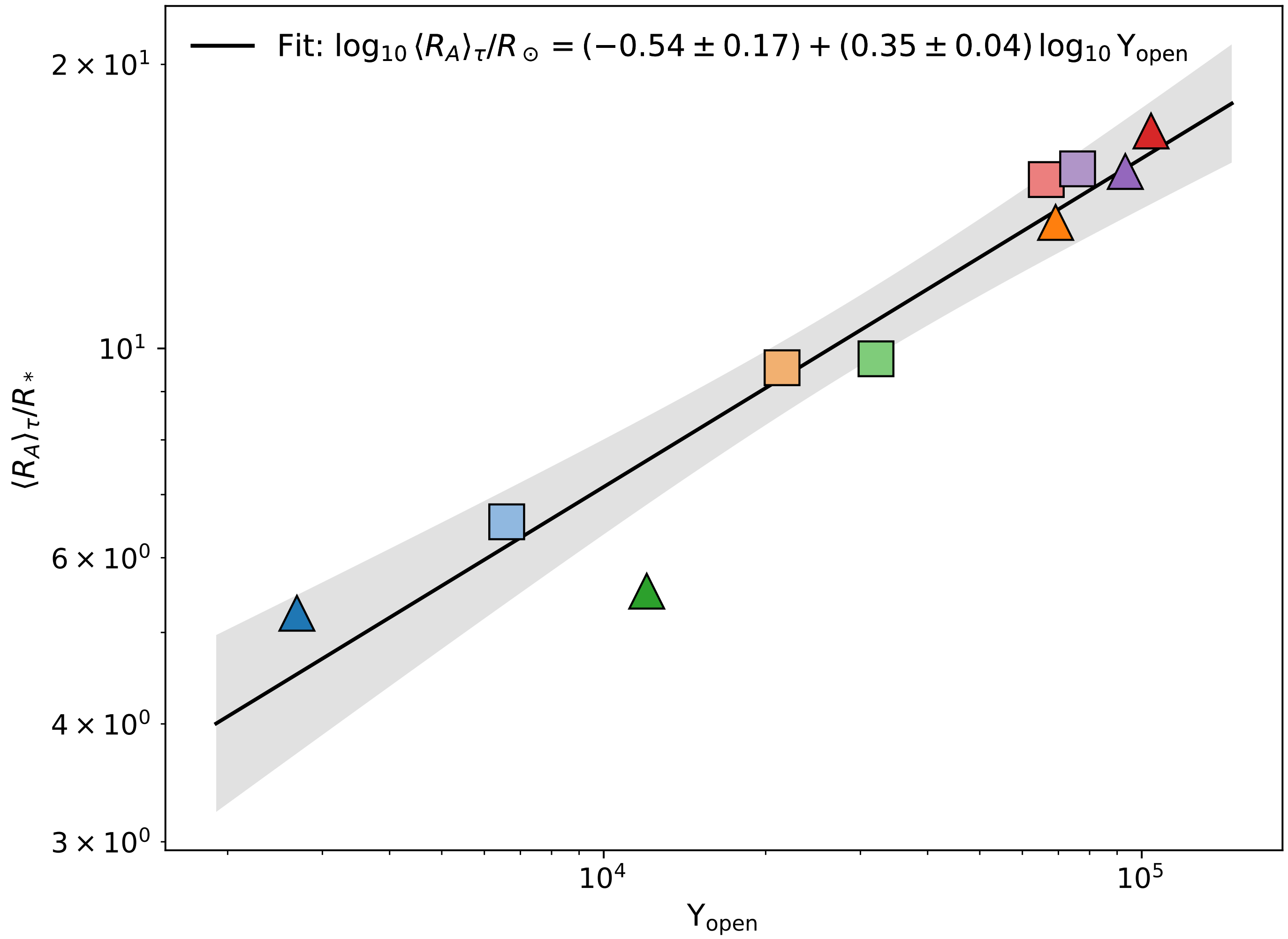

**Figure 6.** Relation between $\Upsilon_{\rm open}$ and $\langle R_{\rm A}\rangle_\tau/R_*$. The blue, orange, green, red, and purple colors correspond to the Sun and C1–C4 cases, respectively. The black solid line shows the best-fit power law in log–log scale, and the shaded gray region indicates the 95% confidence of the fit. $R^2 = 0.92$.

Assuming an Earth-like planet is located in the HZ, we estimate its magnetopause stand-off distance $r_M$, which provides a physically intuitive measure of how strongly the stellar wind compresses a planetary magnetosphere. The pressure balance at the magnetopause is expressed as:

$$p_{\rm w} \simeq \frac{(aB_{\rm p,r_M})^2}{8\pi K}, \tag{8}$$

in Gaussian-cgs units and a dipolar planetary field, $B_{\rm p,r_M} = B_p\left(\frac{R_p}{r_M}\right)^3$, where $B_p = 0.31$ G is the surface equatorial field and $R_p$ is the planetary radius. We adopt the standard pressure balance at the subsolar magnetopause (G. D. Mead 1964; G. D. Mead & D. B. Beard 1964), with $a = 2.44$ for the self-consistent Chapman–Ferraro magnetosphere and a pressure coefficient $K \approx 0.88$ from the gas-dynamic solution (J. R. Spreiter et al. 1966). The results are also listed in Tables 1 and 2.

Our models get 9–10 $R_p$ for solar cases, consistent with the typical observational value of $\sim 10\,R_p$ (J.-H. Shue et al. 1997). More active solar-type stars produce markedly smaller $r_M$, i.e., a more strongly compressed magnetosphere. In the HZ of C1–C4 cases, the orbit-median $r_M$ can fall below $\sim 4\,R_p$. A smaller stand-off distance indicates a more easily compressed magnetosphere; under favorable conditions, it may also be more prone to cause dayside reconnection with the interplanetary magnetic field, increasing the open flux. We note that, although the link between magnetospheric size and habitability itself remains debated (R. Ramstad & S. Barabash 2021), $r_M$ nonetheless provides a direct and physically meaningful diagnostic of the stellar-wind forcing experienced by the planets.

In the foregoing analysis, we characterize the wind environment along a single orbital track, thereby representing a median description of the inherently 3D structures of $p_{\rm w}$. Nevertheless, these 3D effects are particularly worth examining in scenarios involving spin–orbit misaligned exoplanets (J. N. Winn & D. C. Fabrycky 2015). In particular, from C2 to C4 the $p_{\rm w}$ values are mostly of the same magnitude, but the global morphology becomes progressively more “wrapped” and spiral-like as $\Omega$ increases. The wind structure is increasingly shaped by rotational forcing, which can become as important as the effect of magnetic field topology.

## 4. Discussions

### 4.1. Mass-loss Rate–X-Ray Flux Relation

In Section 3.2.2, our results show a clear relation $\dot{M}/R_*^2 \propto F_X^{0.67}$. The element we used to calibrate against observations is the coronal heating prescription, tuned so that the predicted X-ray luminosities fall within the observed range at a given activity level (see Paper I). This calibration fixes the normalization of $F_X$, but leaves the scaling of $\dot{M}/R_*^2$–$F_X$ unconstrained. Therefore, the slope of the $\dot{M}/R_*^2$–$F_X$ relation emerges as a genuine prediction, arising from the global Alfvén-wave driving solution of the wind over the dynamo-generated surface fields. A derivation of this relation from first principles would require chaining multiple intermediate power-law scalings (for example, from $\dot{M}$ to relevant magnetic characteristics, then from those magnetic characteristics to associated energy input, and finally from the energy input to resulting X-ray emission). The cumulative propagation of uncertainties across these steps makes such an approach unreliable. Instead, we approach this by examining their underlying magnetic drivers. In our models, the steady-wind mass-loss rate is primarily controlled by the open magnetic flux (and hence by the large-scale component of the surface field; see Appendix A), whereas the $F_X$ traces the confined plasma and hence is dominated by the closed fields. Consequently, a physical correlation between $\dot{M}$ and $F_X$ necessitates that the open and closed magnetic components scale together. To test this correlation observationally, we rely on empirical proxies motivated by a simple physical picture. Open field lines extend far from the stellar surface, where small-scale multipoles have already decayed; the open flux is therefore dominated by the large-scale field, which is precisely the component recovered by ZDI as $\langle B_V \rangle$. Closed coronal structures, in contrast, are determined by magnetic structures on all scales, so the unsigned field strength $\langle B_I \rangle$ inferred from Stokes-$I$ Zeeman diagnostics provides a natural proxy for it.[10] Empirical studies confirm a strong link between these components. V. See et al. (2019b) find $\langle B_I \rangle \propto \langle B_V \rangle^{0.70\pm0.06}$ across F to M dwarfs ($\langle B_I \rangle \propto \langle B_V \rangle^{0.78\pm0.12}$ for $M_\star > 0.5\,M_\odot$), while the homogeneous G-dwarf analysis by O. Kochukhov et al. (2020) reports a flatter relation of $\langle B \rangle \propto \langle B_V \rangle^{0.48\pm0.05}$. The empirical $\langle B_V \rangle$–$\langle B_I \rangle$ relation shows that multiscale components of stellar surface magnetism do not vary independently, but coevolve as activity changes (see more discussions in V. See et al. 2019b; O. Kochukhov et al. 2020). Ultimately, this multiscale magnetic coupling provides a robust observational foundation for the correlation between $\dot{M}$ and $F_X$ in our models.

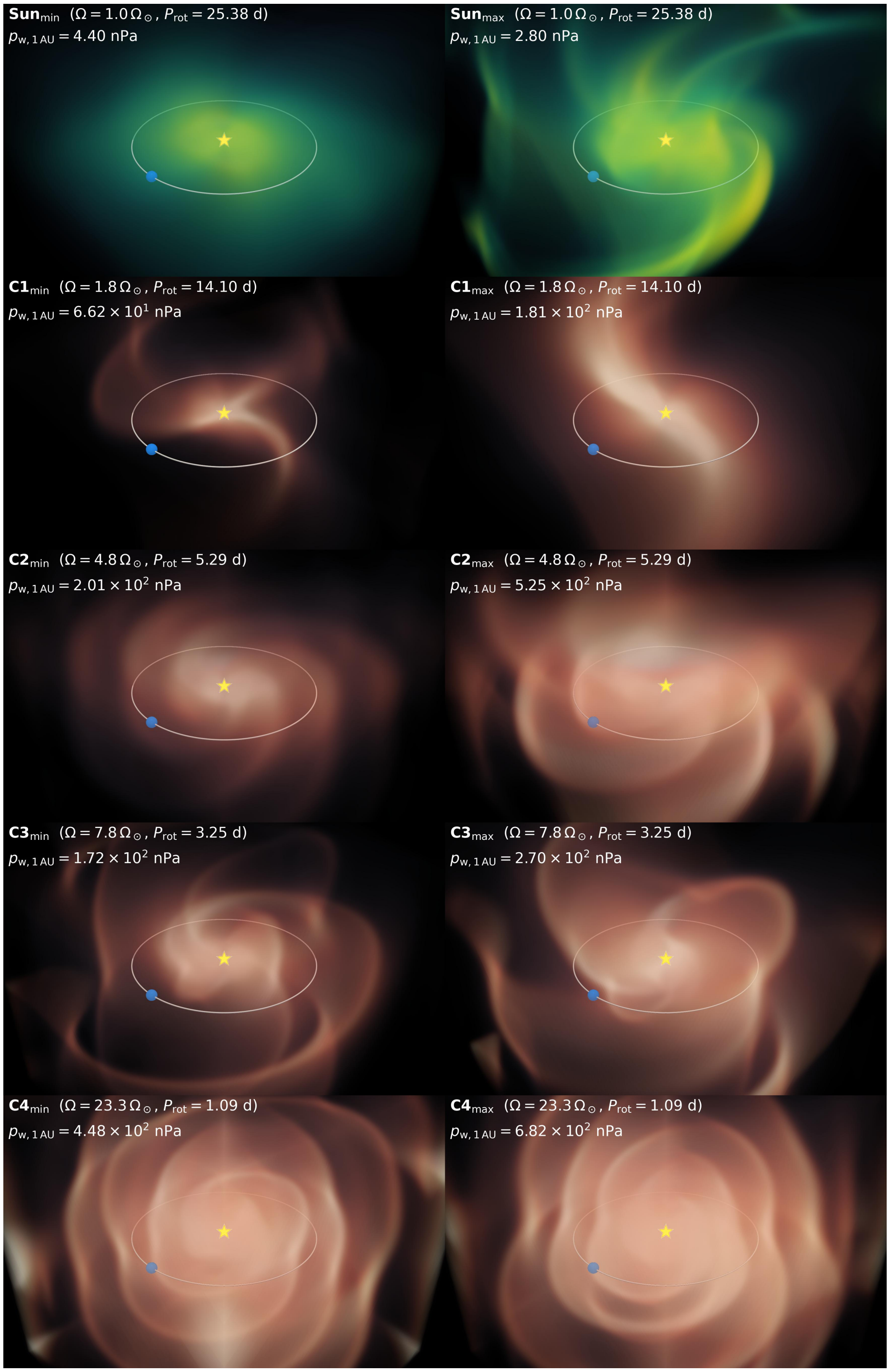

**Figure 7.** Direct volume rendering of the 3D wind pressure. The central star symbol marks the stellar position. The Earth's orbit is shown as a gray ellipse, with a blue dot indicating the Earth. The scene is viewed from a direction inclined by $60\circ$ to the ecliptic plane. Annotations report each star's name, rotation speed, rotation period, and median wind pressure along the Earth's orbit (1 au).

Two caveats deserve emphasis here. First, the surface magnetic fields themselves are taken from global dynamo simulations and inherit their uncertainties, particularly in the partition of multiscale magnetic components. For a more detailed discussion of how modeling outcomes compare with observations and theory in terms of differential rotation and dynamo solutions, see M. Viviani et al. (2018), where most uncertainties are found in the solar-rotation regime, while better agreement is obtained for rapid rotators. However, M. Viviani et al. (2018) did not address the partitioning between small and large scales because simulations at that time did not reach resolutions that allowed for small-scale dynamo (SSD) action. Such models are now available (J. Warnecke et al. 2025), and they indicate that SSD action can further intensify fluctuations (see also E. Işık et al. 2026). Hence, results that are sensitive to these fluctuations may still change with increasing numerical resolution. We therefore regard a slope rather than a very precise value of 0.67 as the most meaningful result of this comparison. Second, our model does not reproduce the high-activity "wind dividing line," at which the empirical $\dot{M}/R_*^2$–$F_X$ relation appears to possibly break down for the most active stars ($F_X \gtrsim 10^6$ erg s$^{-1}$ cm$^{-2}$; B. E. Wood et al. 2005, 2021); in our simulations $\dot{M}$ continues to increase monotonically with activity. More stringent tests will require progress on both the modeling and observational sides. On the modeling side, this means independent constraints on the open-to-closed flux ratio, or coupling our wind framework to dynamo models with better-characterized small-scale magnetic fields. On the observational side, the empirical relation carries substantial scatter, as shown in Figure 5, which limits how models can be tested against the data at present. Also, these astrospheric detections should account for a possible contribution from transient ejecta (e.g., coronal mass ejections; S. R. Cranmer 2017; D. Ó Fionnagáin et al. 2022) and, for very rapid rotators, possibly slingshot-prominence material (M. Jardine & A. Collier Cameron 2019). Expanding the sample of stars with reliable $\dot{M}$ and pairing them with simultaneous X-ray detection are needed in the future.

### 4.2. Magnetic Braking Law

In Section 3.2.3, we quantify the relation between the dimensionless torque-defined effective Alfvén radius $\frac{\langle R_A \rangle_\tau}{R_*}$ and the dimensionless magnetization parameter $\Upsilon_{\rm open}$. In our fitting, Equation (6) degenerates to the form that seems to neglect the rotation factor: $\langle R_A \rangle_\tau / R_\star \simeq 0.29\,\Upsilon_{\rm open}^{0.35}$.

*Exponent.* Our power-law index (0.35) matches that of A. J. Finley & S. P. Matt (2018; polytropic, 2.5D; $m = 0.33$) and M. Shoda et al. (2020; Alfvén-wave-driven, 1D; $m = 0.36$), and is only slightly higher than V. Réville et al. (2015; polytropic, 2.5D; 0.31). It suggests that this braking law scaling is a feature of magnetized stellar winds largely insensitive to dimensionality and reasonable heating prescription.

*Normalization.* Our prefactor (0.29) is smaller than those of V. Réville et al. (2015; 0.64), M. Shoda et al. (2020; 0.38), and A. J. Finley & S. P. Matt (2018; 0.46). Importantly, this suggests that our 3D model predicts a spin-down torque lower than that inferred from 1D or 2.5D models at a given $\Upsilon_{\rm open}$.

*Rotation parameter f.* The degeneracy implies that our results do not require an explicit dependence on the rotation parameter $f$. In the present work, all simulated cases have $f \lesssim 0.11$, placing them squarely on the slow-rotator plateau of the S. P. Matt et al. (2012) formula. Thus, the parameter $f$ is expected to have a negligible influence on the ratio $r_A/R_*$ for stars as slowly rotating as those considered here.

We also find an outlier C2max, which deviates significantly from the best-fit scaling. Although its magnetic strength is comparable to the C3max, it gets a markedly different wind. We suspect that the difference in the magnetic geometry may be responsible for their variance. Actually, the magnetic geometry in C2max appears to have more closed magnetic configurations, in which a substantial fraction of the flux closes at low heights, and the open channels become fewer. It reduces $\Phi_{\rm open}$, causes the Alfvén surface to contract inward, and shortens the effective lever arm. As a result, the magnetic field exerts weaker control over angular-momentum exchange. Meanwhile, the rather large $\dot{M}$ (in Figure 4, C2max's $\dot{M}$ follows the overall trend) indicates that the wind remains strongly mass loaded, and the magnetic field needs to enforce it at a smaller radius, which is harder. We decomposed the angular-momentum-loss rate into a matter term $\rho v_r v_\phi$ and a Maxwell-stress term $-B_r B_\phi/4\pi$,[11] and find that the negative contribution is primarily from the matter term, consistent with the picture discussed above. By construction, $\Upsilon_{\rm open} \propto \Phi_{\rm open}^2/\dot{M}$ already encodes the leading-order geometric information through the open flux. Therefore, the discrepancy of C2max points to a residual dependence that $\Upsilon_{\rm open}$ cannot capture. The integrand of $\dot{J}_z$ depends not only on the magnitudes of $\rho$, $\mathbf{v}$, and

[10] We note that ZDI recovers only a fraction of the true large-scale field, with the missing fraction depending on the truncation degree $\ell$, $v \sin i$, and inclination (e.g., L. T. Lehmann et al. 2019; T. Hackman et al. 2024). Nevertheless, $\langle B_V \rangle$ and $\langle B_I \rangle$ remain the best available proxies for testing the multiscale magnetic correlation.

[11] For a spherical surface ($\mathbf{n} = \hat{\mathbf{e}}_r$), the torque about the rotation axis can be written as $\dot{J}_z = \oint r \sin\theta\, T_{r\phi}\, dA = \oint r \sin\theta \left( \rho v_r v_\phi - \frac{B_r B_\phi}{4\pi} \right) dA$, where the thermal and magnetic pressure terms do not contribute because they have no $r$ and $\phi$ components.

$\boldsymbol{B}$, but also on the local alignment between $v_r$ and $v_\phi$, and between $B_r$ and $B_\phi$, as well as on how opposite-sign regions cancel over the Alfvén surface. Angular information and 3D information are inevitably averaged out when the wind is converted into a single scalar. A full characterization of the relevant 3D diagnostic and whether such outliers are a common group is left to future work.

### 4.3. Rotational Dependence of Angular-momentum Loss

We verify that the Sun and C1–C3 lie clearly on the unsaturated branch: their surface-averaged magnetic field strengths scale with the Rossby number in a manner consistent with the observed distribution of unsaturated solar-type stars (see Figure 1 of Paper I), and their Rossby numbers $Ro \gtrsim 0.1$ are well above the saturation threshold (e.g., N. J. Wright et al. 2011). C4, the most rapidly rotating model, lies closer to the saturation boundary; however, its magnetic field strength has not yet reached the fully saturated level, allowing it to be reasonably grouped with the unsaturated models. While these five models do not necessarily represent a single star's evolutionary track, they provide representative snapshots across the unsaturated regime. Because our models are physically representative of this domain, the derived torque–rotation relation $\dot{J} \propto \Omega^{2.74\pm0.97}$ is not merely a numerical coincidence. Indeed, this scaling is in agreement with the theoretical expectation for unsaturated solar-type stars ($\dot{J} \propto \Omega^3$) required to naturally recover the classical Skumanich spin-down law ($\Omega \propto t^{-1/2}$) (see also S. D. Kawaler 1988).

Notably, this agreement is achieved using the steady stellar wind alone, without invoking any additional contribution from coronal mass ejections, whose role in angular-momentum loss remains debated (e.g., A. N. Aarnio et al. 2012; Y. Xu et al. 2024). This suggests that, at least for rotating solar-type stars on the unsaturated branch, steady winds may be sufficient to account for the observed spin-down rate.

### 4.4. Application of Wind-pressure Model to Real Planetary Systems

Beyond high-energy radiation, the stellar wind itself constitutes a fundamental driver of planetary atmospheric escape. Decades of solar system observations have demonstrated that the wind environment influences both the configuration of planetary magnetospheres and the efficiency of escape processes (e.g., R. Ramstad & S. Barabash 2021). At Earth, the $O^+$ outflow rate is found to increase systematically with solar wind pressure (e.g., A. Schillings et al. 2019). Although such couplings are convolved with the effect of X-ray and ultraviolet heating, the particle flux and embedded magnetic field of the wind exert an influence in their own right: stellar-wind-driven processes (including ion pick-up, sputtering, magnetospheric escape, and so on) remain significant contributors to mass loss even in regimes where thermal escape dominates (G. Gronoff et al. 2020). Quantifying the wind conditions incident on the planet is therefore indispensable for any realistic estimate of planetary atmospheric escape.

It is instructive to apply the models to a few representative systems. Using the NASA Exoplanet Archive,[12] we select a sample of published-confirmed super-Earths with $P_{\rm orb} > 100$ days and $R_p \in [0.8, 2.0]$ (in Earth radius), orbiting solar-type stars with $M_\star$, $R_\star \in [0.8, 1.2]$ (in solar units), ages younger than the Sun, and rotation periods shorter than 30 days. Only two exoplanets, Kepler-1185 b and Kepler-1638 b, meet the selection criteria. An additional exoplanet, Kepler-452 b, despite the lack of a measured stellar rotation period, is added to our sample, because it is in the most well-known Sun–Earth analog systems in terms of stellar properties and planetary orbital properties. We list their stellar and planetary properties in Table 3, with the Sun–Earth reference.

By comparing stellar ages and rotation rates, we associate each system with the most appropriate wind scenario in our runs. Specifically, we adopt the solar simulation for Kepler-1638 b and Kepler-452 b, and the C1 case for Kepler-1185 b. Since the spin–orbit (mis)alignment angles of these systems are not available in the literature, we assume that the planets orbit in the stellar equatorial plane. Here we focus exclusively on characterizing the external wind environment, quantified by the stellar-wind pressure $p_{\rm w}$ at the planetary orbit, which greatly impacts the evolution of the planetary atmosphere (R. Ramstad & S. Barabash 2021). The median, minimum, and maximum values of $p_{\rm w}$ are summarized in Table 3.

Kepler-1185 b ($a = 0.43$ au) orbits close to its relatively active, solar-type host star and lies outside the HZ. However, it remains an instructive case for studying close-in super-Earths, or analogs to a young Mercury ($a = 0.39$ au). The orbit-median stellar-wind pressure is ~400–900 nPa, i.e., several hundred times that at Earth's orbit. Due to the 3D structure, the wind imposes a strongly phase-variable pressure on the planetary magnetosphere as the planet alternately crosses dense spiral arms and the more tenuous interarm regions. Within a spiral arm, the wind pressure can reach $P_{\rm w} \sim 1000$–$5000$ nPa, sufficient to collapse the magnetosphere, whereas in the interarm regions it drops to $P_{\rm w} \sim 200$–$800$ nPa and the magnetosphere recovers. The planet, therefore, experiences a periodic compression–relaxation cycle, which drives influence on the exoplanet atmosphere in a manner analogous to repeated impacts of coronal mass ejections. Kepler-1638 b lies near the inner edge of the HZ, near the Recent Venus limit, and experiences an orbit-median wind pressure of order 6–8 nPa, modestly higher than the contemporary Earth value. Kepler-452 b encounters a stellar-wind environment broadly comparable to that of the Earth, consistent with its larger orbital distance and relatively Sun-like host star.

## 5. Conclusions

In this work, by inputting global convective dynamo-generated surface magnetic maps and solar magnetograms into SWMF/AWSoM, we model Alfvén-wave-heated winds for four solar-type stars and the Sun with multiscale magnetic components. The rotation rates span from 1.0 $\Omega_\odot$ to 23.3 $\Omega_\odot$ ($P_{\rm rot}$ of 25.38–1.09 days) and magnetic field strength ranges from 6.0 to 1200 G. Together with our Paper I on stellar coronae, we clarify the distinct roles of multiscale magnetic field structures and connect steady-wind properties to coronal activity within a single, physically consistent modeling framework.

Across our model sequence, faster rotation produces a more tightly wound spiral, a more extended Alfvén surface, a higher-speed terminal wind, and a high-pressure wind environment, which suggests that young and active solar-type

[12] https://exoplanetarchive.ipac.caltech.edu/

**Table 3**
Properties of Planetary Systems in the Habitable Zone, Including the Sun–Earth Reference

| Host Star (Planet) | Kepler-1185 (b) | Kepler-1638 (b) | Kepler-452 (b) | Sun (Earth) |
|---|---|---|---|---|
| Stellar Properties | | | | |
| $M_*$ ($M_\odot$) | $0.96^{+0.02}_{-0.024}$ | $0.817^{+0.092}_{-0.048}$ | $1.037^{+0.054}_{-0.047}$ | 1.0 |
| $R_*$ ($R_\odot$) | $0.87^{+0.02}_{-0.018}$ | $0.835^{+0.189}_{-0.075}$ | $1.11^{+0.15}_{-0.09}$ | 1.0 |
| $T_{\rm eff}$ (K) | $5622^{+68}_{-59}$ | $5906^{+183}_{-147}$ | $5757^{+85}_{-85}$ | 5772 |
| $P_{\rm rot}$ (d) | 12.81 | 27.7 | … | ∼25 |
| Age (Gyr) | $1.55^{+0.69}_{-0.37}$ | $4.37^{+4.17}_{-2.59}$ | $3.55^{+2.70}_{-1.84}$ ($6^{+2}_{-2}$) | 4.57 |
| Planetary Properties | | | | |
| $R_{\rm p}$ ($R_\oplus$) | $1.33^{+0.08}_{-0.05}$ | $1.87^{+0.32}_{-0.19}$ | $1.63^{+0.23}_{-0.20}$ | 1.0 |
| $M_{\rm p}$ ($M_\oplus$) | 2.33 | 4.16 | 3.29 | 1.0 |
| $P_{\rm orbit}$ (d) | $104^{+0.0007}_{-0.0007}$ | $259^{+0.014}_{-0.014}$ | $385^{+0.007}_{-0.012}$ | 365.25 |
| $a$ (AU) | 0.43 | 0.75 | 1.05 | 1.0 |
| $S$ ($S_\oplus$) | … | $1.17^{+0.58}_{-0.28}$ | $1.10^{+0.29}_{-0.22}$ | 1.0 |
| $T_{\rm eq}$ (K) | … | $304^{+39}_{-31}$ | $265^{+15}_{-13}$ | ∼255 |
| Stellar-wind and Magnetospheric Quantities | | | | |
| $p_{\rm w,min}$ (nPa) | $4.12 \times 10^2$ [$2.36 \times 10^2$, $1.60 \times 10^3$] | 8.02 [6.17, 9.68] | 3.99 [3.12, 5.16] | 4.40 [3.42, 5.63] |
| $p_{\rm w,max}$ (nPa) | $9.12 \times 10^2$ [$7.39 \times 10^2$, $4.86 \times 10^3$] | 5.90 [2.84, $9.35 \times 10^1$] | 2.47 [1.14, $4.12 \times 10^1$] | 2.80 [1.31, $4.59 \times 10^1$] |

**Notes.** Parameter definitions. $M_*$: stellar mass (in solar mass $M_\odot$). $R_*$: stellar radius (in solar radius $R_\odot$). $T_{\rm eff}$: stellar effective temperature (K). $P_{\rm rot}$: stellar rotation period (days). Age: stellar age (Gyr). $R_{\rm p}$: planetary radius (in Earth radius $R_\oplus$). $M_{\rm p}$: planetary mass (in Earth mass $M_\oplus$). $P_{\rm orbit}$: orbital period (days). $a$: orbital semimajor axis (au). $S$: orbit-averaged irradiance in units of the present-day Earth value $S_\oplus$. $T_{\rm eq}$: planetary equilibrium temperature (K). $p_{\rm w}$: median stellar-wind pressure at the planet's orbital location (nPa). Values in square brackets are given as [orbital min, orbital max]. Subscripts "min" and "max" attached to wind/magnetospheric quantities denote stellar activity minimum and maximum, respectively. Planet masses $M_{\rm p}$ and semimajor axes $a$ are taken from the NASA Exoplanet Catalog (https://science.nasa.gov/exoplanets/exoplanet-catalog/). Solar/Earth constants follow IAU 2015 Resolution B3 (A. Prša et al. 2016). The stellar rotation periods of Kepler-1185 and Kepler-1638 are from T. Reinhold et al. (2013) and T. Mazeh et al. (2015), respectively.

stars produce wind environments markedly different from that of the present-day Sun.

We also compute the mass-loss rate $\dot{M}$ and angular-momentum-loss rate $\dot{J}$. It is found that our results differ systematically from previous predictions, most notably from other AWSoM-based studies based on ZDI-reconstructed magnetic maps. This arises from (i) the fact that they utilized ZDI-reconstructed maps that tend to underestimate the true field strength; and (ii) a solar-calibrated Alfvén-wave Poynting-flux input, whereas we scale the injected wave energy self-consistently with $L_X$ (see Paper I). We suggest that reliable stellar-wind modeling requires an accurate surface magnetic field and a physically motivated prescription for energy input.

Based on the successful reproduction of X-ray coronae in our Paper I, we compare our simulated $\dot{M}$ with astrospheric Ly$\alpha$ constraints from observations, and yield a $\dot{M} \propto F_X^{0.67}$ scaling, which is close to the empirical scaling derived from observations ($\dot{M} \propto F_X^{0.77}$; B. E. Wood et al. 2021). This agreement is recovered only when small-scale magnetic components are included in the surface field input (here via the dynamo-generated maps), highlighting the importance of multiscale magnetic components for linking coronal X-ray activity to wind mass loss. The break of $\dot{M}$–$F_X$ scaling with more active stars does not emerge naturally from the steady-wind physics included here over the parameter range we explored. We further quantify magnetic braking in terms of the open-flux magnetization, obtaining $\frac{\langle R_{\rm A}\rangle_\tau}{R_*} = 0.29\Upsilon_{\rm open}^{0.35}$. We also obtain an empirical fit $\dot{J} \propto \Omega^{2.76\pm0.97}$, which closely matches the $\sim\Omega^3$ empirically required by the classical Skumanich law. It indicates that steady stellar winds alone, without the need to invoke coronal mass ejections, are sufficient to drive the substantial angular-momentum loss observed in main-sequence solar-type stars.

Finally, we quantify the wind pressure throughout the 3D space and assess its impact on space weather. Applying our wind solutions to three representative super-Earths (Kepler-1185 b, Kepler-1638 b, and Kepler-452 b; selected from the NASA Exoplanet Archive), we find that close-in planets (Kepler-1185 b) around a younger solar-type star can experience orbit-median wind pressures hundreds of times larger than at Earth. Along its orbit, the assumed Earth-like magnetosphere may experience periodic collapse and relaxation. For those whose host stars are almost at the same age as the Sun, planets near the inner HZ edge (Kepler-1638 b) face only modest enhancements, and a Sun–Earth analog experiences a broadly Earth-like environment (Kepler-452 b). Beyond high-energy radiation, the stellar wind is comparatively less well characterized, yet it is an environmental forcing that, while not always dominant, is essential for certain escape channels such as ion pick-up, sputtering, and magnetospheric escape (G. Gronoff et al. 2020; R. Ramstad & S. Barabash 2021). Future work should couple the stellar-wind environment with radiative influence (e.g., D. A. Brain et al. 2026).

Future work might also consider other drivers of the interplanetary environment besides stellar winds. Transient phenomena such as coronal mass ejections (and possibly prominence eruptions in fast rotators) can also contribute to stellar mass and angular-momentum loss, and their signatures in the interplanetary medium may be difficult to distinguish from the background winds. Moreover, eruptions and associated high-energy particle events can strongly perturb planetary habitability. A more complete assessment will therefore require future work incorporating these transient and energetic processes.

## Acknowledgments

The numerical calculations in this paper were done on the computing facilities in the High Performance Computing Center of Nanjing University. SDO is a mission of NASA's Living With a Star (LWS) program. Hinode is a Japanese mission developed and launched by ISAS/JAXA, with NAOJ as a domestic partner and NASA and STFC (UK) as international partners. It is operated by these agencies in cooperation with ESA and NSC (Norway). CHIANTI is a collaborative project involving George Mason University, the University of Michigan (USA), University of Cambridge (UK), and NASA Goddard Space Flight Center (USA). This work is supported by the National Natural Science Foundation of China under grants 12525305 and 12403066, the Fundamental Research Funds for the Central Universities (KG202506), the Jiangsu Funding Program for Excellent Postdoctoral Talent, the Postgraduate Research & Practice Innovation Program of Jiangsu Province (Project No. KYCX24_0182), and the Program of China Scholarship Council. V.S. acknowledges support from the European Research Council (ERC) under the European Union's Horizon 2020 Framework Programme (CartographY G.A. No. 804752).

## Appendix A
## Impact of Small-scale Magnetic Structures on Wind Modeling

As introduced in Section 1, most 3D wind models are recently driven either by idealized analytic magnetic fields (e.g., dipolar, quadrupolar configurations, etc.) or by observed ZDI maps. The main difference between those studies and ours is that we adopt dynamo-generated magnetic maps that include substantial small-scale structure. Actually, in our model, the average magnetic field strength is also used to scale the injected Poynting flux. Although we have shown in Paper I that small-scale fields are important in coronal heating and reproducing coronal properties, it remains unclear to what extent they influence the global wind solution.

To control variables as much as possible, we perform an experiment based on C3$_{\rm max}$: all parameters are kept fixed, and we only truncate the spherical-harmonic coefficient of the surface map to $l_{\rm max} = 5$. The input magnetic map and the resulting wind structure are shown in Figure A1. The key quantitative results for this C3$_{\rm max}^{\rm LR}$ case are listed in Table 2. The "LR" means the magnetic map is low resolution, but the simulation resolution keeps the same as that in C3$_{\rm max}$.

The mean field strength at the inner boundary differs (875 G in C3$_{\rm max}$ versus 428 G in C3$_{\rm max}^{\rm LR}$). The stellar-wind mass-loss rate, which can be sensitive to coronal conditions, shows a decrease from $1.43 \times 10^{14}$ g s$^{-1}$ in C3$_{\rm max}$ to $1.30 \times 10^{14}$ g s$^{-1}$ in C3$_{\rm max}^{\rm LR}$. By contrast, the open magnetic flux, $\Phi_{\rm open}$ ($\simeq 2.12 \times 10^{24}$ Mx), and the angular-momentum-loss rate, $\dot{J}$ ($\simeq 4.44 \times 10^{33}$ erg), do not change significantly. As a result, the torque-defined effective Alfvén radius varies only slightly, from $\langle R_{\rm A} \rangle_\tau = 17.0\, R_*$ in C3$_{\rm max}$ to 17.8 $R_*$ in C3$_{\rm max}^{\rm LR}$. The geometric-mean Alfvén surface radius, $\langle R_{\rm A} \rangle_{\rm geo}$, shows a similarly small increase (from 27.3 $R_*$ to 28.3 $R_*$). The wind-speed distribution and its probability density function are also very similar between the two runs.

Overall, these results indicate that, for the purpose of global stellar wind modeling, low-resolution magnetic maps can yield wind solutions comparable to those obtained with much higher-resolution maps, at least for the cases considered here. However, although spherical-harmonic truncation produces a ZDI-like large-scale morphology, it is not completely equivalent to a ZDI map, since it applies an explicit low-pass filter rather than the cancellation- and resolution-limited reconstruction in ZDI. Therefore, when comparing winds driven by different magnetic maps, matching the overall field amplitude or an appropriate proxy such as unsigned flux should remain a primary consideration.

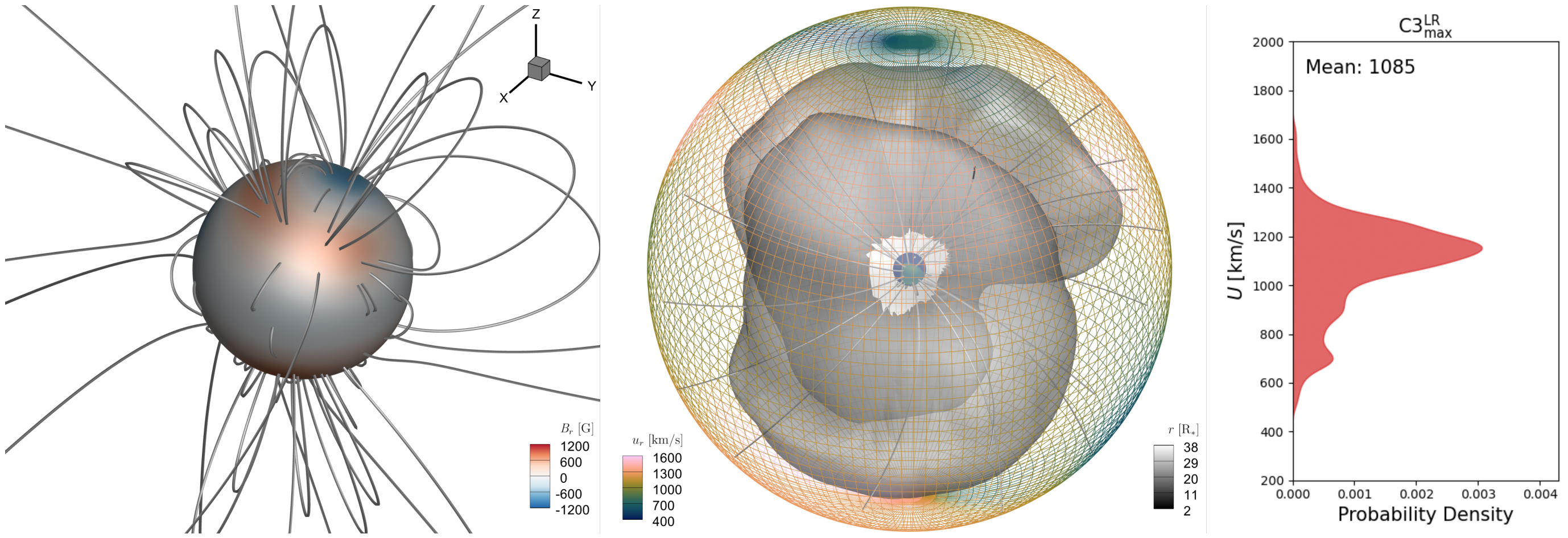

**Figure A1.** The stellar-wind solution for C3$_{\rm max}^{\rm LR}$. Left: input radial magnetic field map truncated at $l_{\rm max} = 5$. Middle: three-dimensional wind structure shown in the same style as Figure 2. Right: probability density distribution of the wind speed measured at $r = 36\, R_*$.

While our truncation test indicates that small-scale magnetic structure has only a minor impact on the global solution, this conclusion is based on global proxies, i.e., we compress the information of 3D winds into several global values like average Alfvén radius and loss rates, and do not preclude more localized differences. That does not mean localized effects can be neglected. In particular, the expansion of magnetic flux tubes above the stellar surface (or the fraction of the surface covered by magnetic regions) can influence wind dynamics (Y.-M. Wang & N. R. Sheeley 1990; Z. Wu et al. 2026). Moreover, fine magnetic structures can modulate the wind-speed distribution (e.g., through topological complexity akin to the solar Separatrix-Web scenario by A. K. Higginson & B. J. Lynch 2018 and B. J. Lynch et al. 2023), but exploring such effects is beyond the scope of the present study.

## Appendix B
## Statistics of the Coronal State

Parameterized wind models (e.g., O. Cohen & J. J. Drake 2014) often investigate the impact of an assumed "coronal base density" setting, which is explicitly defined as the inner-boundary density imposed in the simulation domain. To get a cross-model benchmarking, we report mean statistics of the coronal state at a fixed height, $r = 1.1\,R_*$ (Table B1). They describe the typical low-coronal conditions. Values in brackets give the P10–P90 range, which quantifies the spatial inhomogeneity on the $r = 1.1\,R_*$ shell. The reliability of the coronal properties has been discussed in Paper I via emission measure and X-ray comparisons.

In Section 3.2.1, we use the density and magnetic field strength estimates described above as inputs to the O. Cohen &

**Table B1**
Summary of Coronal Base Properties for All Cases

| Case | $n$ (cm$^{-3}$) | $T_e$ (K) | $T$ (K) | $\langle\|\boldsymbol{B}\|\rangle$ (G) |
|---|---|---|---|---|
| $\mathrm{Sun}_{\mathrm{min}}$ | $5.04 \times 10^7$ [$2.95 \times 10^7$, $7.38 \times 10^7$] | $1.13 \times 10^6$ [$8.37 \times 10^5$, $1.46 \times 10^6$] | $1.09 \times 10^6$ [$8.08 \times 10^5$, $1.44 \times 10^6$] | 2.42 [$4.67 \times 10^{-1}$, 5.81] |
| $\mathrm{Sun}_{\mathrm{max}}$ | $1.51 \times 10^8$ [$5.15 \times 10^7$, $3.02 \times 10^8$] | $1.78 \times 10^6$ [$1.04 \times 10^6$, $2.64 \times 10^6$] | $1.76 \times 10^6$ [$1.00 \times 10^6$, $2.64 \times 10^6$] | 4.95 [$7.09 \times 10^{-1}$, $1.14 \times 10^1$] |
| $\mathrm{C1}_{\mathrm{min}}$ | $6.09 \times 10^8$ [$1.15 \times 10^8$, $9.52 \times 10^8$] | $5.61 \times 10^6$ [$3.87 \times 10^6$, $7.21 \times 10^6$] | $5.29 \times 10^6$ [$2.56 \times 10^6$, $7.10 \times 10^6$] | $1.24 \times 10^2$ [$3.22 \times 10^1$, $2.36 \times 10^2$] |
| $\mathrm{C1}_{\mathrm{max}}$ | $1.03 \times 10^9$ [$2.53 \times 10^8$, $1.60 \times 10^9$] | $7.98 \times 10^6$ [$5.84 \times 10^6$, $9.77 \times 10^6$] | $7.71 \times 10^6$ [$4.64 \times 10^6$, $9.79 \times 10^6$] | $2.29 \times 10^2$ [$9.29 \times 10^1$, $3.87 \times 10^2$] |
| $\mathrm{C2}_{\mathrm{min}}$ | $7.51 \times 10^8$ [$4.90 \times 10^8$, $9.46 \times 10^8$] | $6.64 \times 10^6$ [$4.75 \times 10^6$, $8.63 \times 10^6$] | $6.35 \times 10^6$ [$4.69 \times 10^6$, $8.42 \times 10^6$] | $2.19 \times 10^2$ [$2.35 \times 10^1$, $4.56 \times 10^2$] |
| $\mathrm{C2}_{\mathrm{max}}$ | $1.15 \times 10^9$ [$2.95 \times 10^8$, $1.78 \times 10^9$] | $8.68 \times 10^6$ [$5.38 \times 10^6$, $1.08 \times 10^7$] | $8.38 \times 10^6$ [$4.56 \times 10^6$, $1.07 \times 10^7$] | $3.09 \times 10^2$ [$8.02 \times 10^1$, $5.41 \times 10^2$] |
| $\mathrm{C3}_{\mathrm{min}}$ | $1.02 \times 10^9$ [$3.68 \times 10^8$, $1.39 \times 10^9$] | $7.45 \times 10^6$ [$5.27 \times 10^6$, $9.48 \times 10^6$] | $7.26 \times 10^6$ [$4.90 \times 10^6$, $9.37 \times 10^6$] | $2.12 \times 10^2$ [$5.71 \times 10^1$, $4.01 \times 10^2$] |
| $\mathrm{C3}_{\mathrm{max}}$ | $1.34 \times 10^9$ [$5.88 \times 10^8$, $1.85 \times 10^9$] | $9.30 \times 10^6$ [$7.01 \times 10^6$, $1.09 \times 10^7$] | $9.10 \times 10^6$ [$6.82 \times 10^6$, $1.09 \times 10^7$] | $3.09 \times 10^2$ [$1.06 \times 10^2$, $4.96 \times 10^2$] |
| $\mathrm{C4}_{\mathrm{min}}$ | $1.58 \times 10^9$ [$6.49 \times 10^8$, $2.20 \times 10^9$] | $1.03 \times 10^7$ [$6.52 \times 10^6$, $1.33 \times 10^7$] | $9.97 \times 10^6$ [$6.39 \times 10^6$, $1.31 \times 10^7$] | $5.33 \times 10^2$ [$1.06 \times 10^2$, $9.85 \times 10^2$] |
| $\mathrm{C4}_{\mathrm{max}}$ | $2.24 \times 10^9$ [$1.14 \times 10^9$, $2.98 \times 10^9$] | $8.62 \times 10^6$ [$5.58 \times 10^6$, $1.09 \times 10^7$] | $1.00 \times 10^7$ [$5.64 \times 10^6$, $1.38 \times 10^7$] | $6.49 \times 10^2$ [$1.67 \times 10^2$, $1.09 \times 10^3$] |

**Note.** We report the mean statistics of the coronal state at a fixed height, $r = 1.1\,R_*$, which describe the typical low-coronal conditions. The tabulated quantities are the electron/proton number density $n$, the electron temperature $T_e$, the proton temperature $T$, and the magnetic field strength $\langle|\boldsymbol{B}|\rangle$. Values in brackets give the P10–P90 range, which quantifies the spatial inhomogeneity on the $r = 1.1\,R_*$ shell.

J. J. Drake (2014) scaling law to obtain independent predictions of $\dot{M}$ and $\dot{J}$ for our four targets (C1–C4), which we then compare against our own model results. The scaling laws are: $\frac{\dot{M}}{\dot{M}_\odot} = K\left(\frac{n}{n_\odot}\right)^{\alpha}\left(\frac{B}{B_\odot}\right)^{\left(\frac{n_\odot}{n}\right)^{\beta}}\left(\frac{P_\odot}{P}\right)^{\left(1-\frac{n_\odot}{n}\right)^{\gamma}}$ and $\frac{\dot{J}}{\dot{J}_\odot} = \left(\frac{P_\odot}{P}\right)\left(\frac{\dot{M}}{\dot{M}_\odot}\right)$, where we adopted the solar normalization constants $\dot{M}_\odot = 3 \times 10^{-14} M_\odot\,\mathrm{yr}^{-1}$, $\dot{J}_\odot = 2 \times 10^{29}\,\mathrm{g\,cm^2\,s^{-2}}$, $n_\odot = 2 \times 10^8\,\mathrm{cm}^{-3}$, $B_\odot = 10$ G, and $P_\odot = 27$ days, together with the best-fit parameters $K = 3$, $\alpha = 0.8$, $\beta = 0.2$, and $\gamma = 0.1$ suggested in their research.

## ORCID iDs

Yue-Hong Chen (陈悦虹) https://orcid.org/0000-0003-1220-1582
Julián D. Alvarado-Gómez https://orcid.org/0000-0001-5052-3473
Xin Cheng https://orcid.org/0000-0003-2837-7136
Victor See https://orcid.org/0000-0001-5986-3423
Yu Dai https://orcid.org/0000-0001-9856-2770
Maarit J. Korpi-Lagg https://orcid.org/0000-0002-9614-2200
Jörn Warnecke https://orcid.org/0000-0002-9292-4600
Chen Xing https://orcid.org/0000-0002-6550-1522
Mingde Ding https://orcid.org/0000-0002-4978-4972

## References

Aarnio, A. N., Matt, S. P., & Stassun, K. G. 2012, ApJ, 760, 9
Airapetian, V. S., Barnes, R., Cohen, O., et al. 2020, IJAsB, 19, 136
Airapetian, V. S., Jin, M., Lüftinger, T., et al. 2021, ApJ, 916, 96
Alvarado-Gómez, J. D., Hussain, G. A. J., Cohen, O., et al. 2016a, A&A, 588, A28
Alvarado-Gómez, J. D., Hussain, G. A. J., Cohen, O., et al. 2016b, A&A, 594, A95
Boro Saikia, S., Lueftinger, T., Airapetian, V. S., et al. 2023, ApJ, 950, 124
Bouvier, J., Matt, S. P., Mohanty, S., et al. 2014, in Protostars and Planets VI, ed. H. Beuther et al. (Univ. Arizona Press), 433
Brain, D. A., Cohen, O., Cravens, T. E., et al. 2026, arXiv:2603.11561
Chebly, J. J., Alvarado-Gómez, J. D., Poppenhäger, K., & Garraffo, C. 2023, MNRAS, 524, 5060
Chen, Y.-H., Alvarado-Gómez, J. D., Cheng, X., et al. 2025, ApJ, 995, 83
Cohen, O. 2011, MNRAS, 417, 2592
Cohen, O., & Drake, J. J. 2014, ApJ, 783, 55
Crameri, F., Shephard, G. E., & Heron, P. J. 2020, NatCo, 11, 5444
Cranmer, S. R. 2017, ApJ, 840, 114
Cranmer, S. R., Chhiber, R., Gilly, C. R., et al. 2023, SoPh, 298, 126
Cranmer, S. R., & Saar, S. H. 2011, ApJ, 741, 54
Evensberget, D., Carter, B. D., Marsden, S. C., Brookshaw, L., & Folsom, C. P. 2021, MNRAS, 506, 2309
Evensberget, D., Carter, B. D., Marsden, S. C., et al. 2022, MNRAS, 510, 5226
Evensberget, D., Marsden, S. C., Carter, B. D., et al. 2023, MNRAS, 524, 2042
Finley, A. J., Hewitt, A. L., Matt, S. P., et al. 2019, ApJ, 885, L30
Finley, A. J., & Matt, S. P. 2017, ApJ, 845, 46
Finley, A. J., & Matt, S. P. 2018, ApJ, 854, 78
Finley, A. J., Matt, S. P., & See, V. 2018, ApJ, 864, 125
Garraffo, C., Drake, J. J., & Cohen, O. 2015, ApJ, 807, L6
Garraffo, C., Drake, J. J., Dotter, A., et al. 2018, ApJ, 862, 90
Gombosi, T. I., Chen, Y., Glocer, A., et al. 2021, JSWSC, 11, 42
Gronoff, G., Arras, P., Baraka, S., et al. 2020, JGRA, 125, e27639
Hackman, T., Kochukhov, O., Viviani, M., et al. 2024, A&A, 682, A156
Higginson, A. K., & Lynch, B. J. 2018, ApJ, 859, 6
Huang, Z., Tóth, G., Sachdeva, N., et al. 2023, ApJ, 946, L47
Işık, E., Solanki, S. K., Krivova, N. A., & Shapiro, A. I. 2026, A&A, 711, A48
Jardine, M., & Collier Cameron, A. 2019, MNRAS, 482, 2853
Johnstone, C. P., Güdel, M., Brott, I., & Lüftinger, T. 2015a, A&A, 577, A28
Johnstone, C. P., Güdel, M., Lüftinger, T., Toth, G., & Brott, I. 2015b, A&A, 577, A27
Kawaler, S. D. 1988, ApJ, 333, 236
King, J. H., & Papitashvili, N. E. 2005, JGRA, 110, A02104
Kislyakova, K. G., Güdel, M., Koutroumpa, D., et al. 2024, NatAs, 8, 596
Kochukhov, O., Hackman, T., Lehtinen, J. J., & Wehrhahn, A. 2020, A&A, 635, A142
Kopparapu, R. K., Ramirez, R., Kasting, J. F., et al. 2013, ApJ, 765, 131
Kopparapu, R. K., Ramirez, R. M., SchottelKotte, J., et al. 2014, ApJ, 787, L29
Lehmann, L. T., Hussain, G. A. J., Jardine, M. M., Mackay, D. H., & Vidotto, A. A. 2019, MNRAS, 483, 5246
Lynch, B. J., Viall, N. M., Higginson, A. K., et al. 2023, ApJ, 949, 14
Matt, S. P., Brun, A. S., Baraffe, I., Bouvier, J., & Chabrier, G. 2015, ApJ, 799, L23
Matt, S. P., MacGregor, K. B., Pinsonneault, M. H., & Greene, T. P. 2012, ApJ, 754, L26
Matt, S., & Pudritz, R. E. 2005, ApJ, 632, L135
Mazeh, T., Perets, H. B., McQuillan, A., & Goldstein, E. S. 2015, ApJ, 801, 3
Mead, G. D. 1964, JGR, 69, 1181
Mead, G. D., & Beard, D. B. 1964, JGR, 69, 1169
Meng, X., van der Holst, B., Tóth, G., & Gombosi, T. I. 2015, MNRAS, 454, 3697
Mestel, L. 1968, MNRAS, 138, 359
Mestel, L., & Selley, C. S. 1970, MNRAS, 149, 197
Ó Fionnagáin, D., Kavanagh, R. D., Vidotto, A. A., et al. 2022, ApJ, 924, 115
Parker, E. N. 1958, ApJ, 128, 664
Pesnell, W. D., Thompson, B. J., & Chamberlin, P. C. 2012, SoPh, 275, 3
Prša, A., Harmanec, P., Torres, G., et al. 2016, AJ, 152, 41
Ramstad, R., & Barabash, S. 2021, SSRv, 217, 36
Reinhold, T., Reiners, A., & Basri, G. 2013, A&A, 560, A4
Réville, V., Brun, A. S., Matt, S. P., Strugarek, A., & Pinto, R. F. 2015, ApJ, 798, 116
Sachdeva, N., Manchester, W. B., IV, Sokolov, I., et al. 2023, ApJ, 952, 117
Sachdeva, N., Tóth, G., Manchester, W. B., et al. 2021, ApJ, 923, 176
Sachdeva, N., van der Holst, B., Manchester, W. B., et al. 2019, ApJ, 887, 83
Schillings, A., Slapak, R., Nilsson, H., et al. 2019, EP&S, 71, 70
Schou, J., Scherrer, P. H., Bush, R. I., et al. 2012, SoPh, 275, 229
See, V., Matt, S. P., Finley, A. J., et al. 2019a, ApJ, 886, 120
See, V., Matt, S. P., Folsom, C. P., et al. 2019b, ApJ, 876, 118
Shoda, M., Suzuki, T. K., Matt, S. P., et al. 2020, ApJ, 896, 123
Shue, J.-H., Chao, J. K., Fu, H. C., et al. 1997, JGR, 102, 9497
Sokolov, I. V., van der Holst, B., Oran, R., et al. 2013, ApJ, 764, 23
Spreiter, J. R., Summers, A. L., & Alksne, A. Y. 1966, P&SS, 14, 223, IN1, 251
Sullivan, B., & Kaszynski, A. 2019, JOSS, 4, 1450
Suzuki, T. K., Imada, S., Kataoka, R., et al. 2013, PASJ, 65, 98
Tóth, G., Sokolov, I. V., Gombosi, T. I., et al. 2005, JGRA, 110, A12226
Tóth, G., van der Holst, B., Sokolov, I. V., et al. 2012, JCoPh, 231, 870
van der Holst, B., Huang, J., Sachdeva, N., et al. 2022, ApJ, 925, 146
van der Holst, B., Manchester, W. B., IV, Frazin, R. A., et al. 2010, ApJ, 725, 1373
van der Holst, B., Sokolov, I. V., Meng, X., et al. 2014, ApJ, 782, 81
Vidotto, A. A. 2021, LRSP, 18, 3
Vidotto, A. A., & Bourrier, V. 2017, MNRAS, 470, 4026
Vidotto, A. A., Gregory, S. G., Jardine, M., et al. 2014, MNRAS, 441, 2361
Vidotto, A. A., Jardine, M., Opher, M., Donati, J. F., & Gombosi, T. I. 2011, MNRAS, 412, 351
Viviani, M., Warnecke, J., Käpylä, M. J., et al. 2018, A&A, 616, A160
Wang, Y.-M. 1998, ASPC, 154, 131
Wang, Y.-M., & Sheeley, N. R., Jr. 1990, ApJ, 355, 726
Warnecke, J., Korpi-Lagg, M. J., Rheinhardt, M., Viviani, M., & Prabhu, A. 2025, A&A, 696, A93
Weber, E. J., & Davis, L., Jr 1967, ApJ, 148, 217
Winn, J. N., & Fabrycky, D. C. 2015, ARA&A, 53, 409
Wood, B. E., Müller, H.-R., Redfield, S., & Edelman, E. 2014, ApJ, 781, L33
Wood, B. E., Müller, H.-R., Redfield, S., et al. 2021, ApJ, 915, 37
Wood, B. E., Müller, H.-R., Zank, G. P., & Linsky, J. L. 2002, ApJ, 574, 412
Wood, B. E., Müller, H.-R., Zank, G. P., Linsky, J. L., & Redfield, S. 2005, ApJ, 628, L143
Wright, N. J., Drake, J. J., Mamajek, E. E., & Henry, G. W. 2011, ApJ, 743, 48
Wright, N. J., Newton, E. R., Williams, P. K. G., Drake, J. J., & Yadav, R. K. 2018, MNRAS, 479, 2351
Wu, Z., He, J., Hou, C., et al. 2026, ApJS, 282, 4
Xu, Y., Alvarado-Gómez, J. D., Tian, H., et al. 2024, ApJ, 971, 153
Zahn, J.-P. 1977, A&A, 57, 383